# The structure of the hydrated electron.

# Part 2. A mixed quantum/classical molecular dynamics (MQC MD) – embedded cluster density functional theory: single-excitation configuration interaction (DFT:CIS) study.

Ilya A. Shkrob * [1]

*Chemistry Division , Argonne National Laboratory,*

*9700 S. Cass Ave, Argonne, IL 60439*

William J. Glover, Ross E. Larsen and Benjamin J. Schwartz, * [2]

*Department of Chemistry and Biochemistry,*

*University of California, Los Angeles, CA 90095-1569*

## Abstract

Adiabatic mixed quantum/classical (MQC) molecular dynamics (MD) simulations were used to generate snapshots of the hydrated electron in liquid water at 300 K. Water cluster anions that include two complete solvation shells centered on the hydrated electron were extracted from the MQC MD simulations and embedded in a roughly 18 Å x 18 Å x 18 Å matrix of fractional point charges designed to represent the rest of the solvent. Density functional theory (DFT) with the Becke-Lee-Yang-Parr functional and single-excitation configuration interaction (CIS) methods were then applied to these embedded clusters. The salient feature of these hybrid DFT(CIS)/MQC MD calculations is significant transfer (ca. 18%) of the excess electron's charge density into the *2p* orbitals of oxygen atoms in OH groups forming the solvation cavity. We used the results of these calculations to examine the structure of the singly occupied and the lower unoccupied molecular orbitals, the density of states, the absorption spectra in the visible and ultraviolet, the hyperfine coupling (hfc) tensors, and the infrared (IR) and Raman spectra of these embedded water cluster anions. The calculated hfc tensors were used to compute the Electron Paramagnetic Resonance (EPR) and Electron Spin Echo Envelope Modulation (ESEEM) spectra for the hydrated electron that compared favorably to the experimental spectra of trapped electrons in alkaline ice. The calculated vibrational spectra of the hydrated electron are consistent with the red-shifted bending and stretching frequencies observed in resonance Raman experiments. In addition to reproducing the visible/near IR absorption spectrum, the hybrid DFT model also accounts for the hydrated electron's 190-nm absorption band in the ultraviolet. Thus, our study suggests that to explain several important experimentally observed properties of the hydrated electron, many-electron effects *must* be accounted for: one-electron models that do not allow for mixing of the excess electron density with the frontier orbitals of the first-shell solvent molecules cannot explain the observed magnetic, vibrational, and electronic spectroscopy of this species. Despite the need for multielectron effects to explain these important properties, the ensemble-averaged radial wavefunctions and energetics of the highest occupied and three lowest unoccupied orbitals of the hydrated electrons in our hybrid model are close to the *s*- and *p*-like states obtained in one-electron models. Thus, one-electron models can provide a remarkably good approximation to the multielectron picture of the hydrated electron for many applications; indeed, the two approaches appear to be complementary.

---

## 1. Introduction.

This paper is the second part of a two-part series on the structure of the hydrated electron, $e_{hyd}^-$. In Part 1, [1] we examined several idealized water clusters that trap the electron in their interior using density functional theory (DFT) and *ab initio* methods. We found that octahedral anions with OH groups pointing towards the center of a spherical solvation cavity (known as Kevan's model of hydrated electron) [2,3] account reasonably well for the observed hyperfine coupling (hfc) tensors estimated for the electron trapped in low-temperature alkaline ices using Electron Paramagnetic Resonance (EPR) [3,4,5] and Electron Spin Echo Envelope Modulation (ESEEM) spectroscopies. [2,3,6] Importantly, this result depends critically on the use of a multielectron model [1,14,15,16] (as opposed to the standard one-electron approach), [7-13] because the multielectron model predicts considerable sharing of the excess electron density between the cavity and the frontier *2p* orbitals of the oxygen atoms in the OH groups forming the cavity. Without this sharing, the magnetic resonance results cannot be explained even qualitatively. [1]

Although the results from Part 1 [1] are suggestive, it is not clear whether the representation of the hydrated electron by such idealized gas-phase clusters is acceptable: the hydrated electron is a dynamic entity that constantly samples configurations of water molecules that vibrate, rotate, and otherwise move around. This means that the 'hydrated electron' is not a rigid structure, but a statistical average over many configurations; [7-11] in this regard, hydrated electrons are different from non-solvent-supported chemical species, so a quantitative description of the $e_{hyd}^-$ within the multielectron approach has to address this inherent variability. In other words, it is impossible to find a single representative water anion structure that corresponds to the statistical average known as the 'hydrated electron' in liquid water. Thus, the next logical step in the development of multielectron models of the $e_{hyd}^-$ is finding a means of building this statistical picture. This is the purpose of the present study.

One path to this goal is Car-Parrinello molecular dynamics (CPMD), and in fact, a CPMD calculation of the $e_{hyd}^-$ recently has been implemented by Boero et al. [17] In the present study, we suggest a different approach that is less computationally demanding but appears to successfully capture the essential physics of the problem. Our approach capitalizes on the historical success of one-electron models of electron solvation in rationalizing the dynamics and energetics of electron solvation. [7-13] We used adiabatic mixed quantum/classical (MQC) molecular dynamics (MD) calculations [9] to generate a dynamical trajectory of the hydrated electron in its ground electronic state, and then we extracted temporally well-separated snapshots from this trajectory. These snapshots became the input for multielectron DFT and single-excitation configuration interaction (CIS) calculations. In these multielectron calculations, we explicitly considered only one or two complete solvation shells for the excess electron; the remaining atoms in the simulated solvent were replaced by point charges, a procedure we refer to as matrix embedding. (We provide a detailed justification for this hybrid embedded cluster approach in section 3, where we show that the electron density is tightly localized in the solvation cavity and on the dangling OH groups forming the cavity). It turns out that a few hundred such snapshots are sufficient to build a robust picture of the excess electron in water. We find that there is significant sharing of spin and charge of the excess electron by $O\ 2p$ orbitals in the first-shell water molecules (a result which was hinted at in both Part 1 [1] and other studies [14,15,16]). Based on this study, we picture the ground state of the 'hydrated electron' as an unusual kind of multimer radical anion [3] in which ~20% of the excess electron is shared among several first-shell water molecules and ~80% of the electron occupies a cavity between molecules.

We also find that the sharing of electron density with the first-shell solvent molecules is not only consistent with most of the known experimental properties of the $e_{hyd}^{-}$ but is in fact necessary to account for some of these properties. In the rest of this paper, we demonstrate how our hybrid approach can account for several experimentally observed features of the $e_{hyd}^{-}$, including: (i) the energetics and the equilibrium optical spectrum of the $e_{hyd}^{-}$ in the visible and ultraviolet (UV); [18] (ii) the EPR and ESEEM spectra of the $e_{hyd}^{-}$; [1-6] and (iii) the vibrational (resonance Raman) spectrum of the $e_{hyd}^{-}$. [19,20] Our study is also able to explain why one-electron MQC models have been successful in explaining some of the aforementioned properties of the $e_{hyd}^{-}$ even though to date, such one-electron models have not been able to account for the mixing into the nearby water frontier molecular orbitals: although the true multielectron picture of the $e_{hyd}^{-}$ is complex, on average, the radial density for the highest occupied (HOMO) and the three lowest unoccupied (LUMO) molecular orbitals resemble the $s$-like and $p$-like orbitals predicted by the one-electron models. [8b,9,11] In fact, for some observables (e.g., the optical spectrum in the visible), the fine details of this orbital structure do not matter, and even crude models (e.g., dielectric continuum [13] and semicontinuum models [12,13]) can provide an adequate general description of the electron wavefunction. For other observables (e.g., the spin density maps provided by EPR and ESEEM spectroscopies and the resonance Raman spectrum), this level of approximation is inadequate because the fine details of the orbital structure do matter, and so an approach that can account for mixing of the excess electron into the water frontier molecular orbitals is needed.

The rest of this paper is organized as follows: The computational details of our MQC MD and DFT (CIS) calculations are given in section 2. In section 3.1, we examine the structure of the singly occupied molecular orbital (SOMO), which is also the HOMO. In section 3.2, we analyze the density of states (DOS) of the electron, examine the three lower unoccupied molecular orbitals, and use the DFT and CIS methods to calculate the absorption spectrum of the $e^-_{hyd}$. In section 3.3, we use our DFT data to obtain hyperfine coupling tensors and simulate matrix EPR and $^2$H ESEEM spectra of $e^-_{hyd}$. In section 3.4, we discuss the vibrational properties of the hydrated electron, and compare to recent resonance Raman experiments. Finally, we summarize our results and offer some concluding remarks in section 4.

In order to reduce the length of this paper, we have placed a significant amount of material (figures with the designator "S" (e.g., Figure 1S)) in the Supplementary Information.

## 2. Computational Details.

### 2.1. Mixed Quantum/Classical Molecular Dynamics (MQC MD) calculations.

The electronic structure calculations described in this paper were performed on clusters of water molecules extracted from a 100-ps adiabatic MQC MD trajectory with a time step of 1 fs. In this trajectory, the water molecules moved classically according to the velocity Verlet algorithm [21] and the single excess quantum mechanical electron was confined to its adiabatic ground state. The 200 water molecules occupied a cubic 18.17 Å x 18.17 Å x 18.17 Å simulation cell and interacted with each other through SPCf (flexible simple-point charge) potentials. [22] The excess electron interacted with the water molecules through a pairwise-additive pseudopotential, $V_p$, [23] and at every time step the ground-state wave function, $\psi$, of the electron was calculated on a 16 x 16 x 16 cubic grid using an iterative-and-block-Lanczos algorithm. [24]

The force exerted by the electron on the water molecules was of the Hellmann-Feynman form, $\boldsymbol{F}_i = -\langle\psi|\nabla_i V_p|\psi\rangle$, where $\boldsymbol{F}_i$ is the force on atom $i$ and $\nabla_i$ denotes a gradient with respect to the spatial coordinates of atom $i$. The average temperature of the system was 296 K with root-mean-square fluctuations of 8 K.

We extracted from this trajectory a sequence of 1000 snapshots of the $e_{hyd}^-$ that were separated by a time interval of $\Delta t = 100$ fs. In each of the extracted snapshots, the coordinates were shifted so that the center of mass of the electron, $X$, was at the origin and minimum image periodic boundary conditions were applied. Water molecules were treated explicitly if the distance $r_{XH}$ between $X$ and one of the protons, $H$, was less than a chosen cutoff radius, $r_{cut}$; the cluster size $n$ is then defined as the number of such water molecules. The remaining "matrix" molecules were replaced [25,26] by point charges (chosen to be the same as in the SPCf model), [22] $Q_H = +0.41$ for hydrogen and $Q_O = -2Q_H$ for oxygen (in the following, such water anion and neutral clusters are referred to as "embedded" ones). With the exception of the IR-Raman simulations, the cutoff radius $r_{cut}$ was chosen to be 4.75 Å, which corresponds to the first two solvation shells around the electron cavity; for the IR-Raman calculations, $r_{cut}$ was set to 3 Å (although for the neutral clusters discussed in section 3.4, where there is no electron center of mass, we based the cutoff, chosen to be 3.5 Å, on the distance from one of the $O$ atoms). Hereafter, we define the first solvation shell as including those water molecules for which at least one of the protons has $r_{XH} < 3$ Å. We then label the protons satisfying this condition by $H_{in}$ ('inside'), and the protons *in the first solvation shell* that do not satisfy this condition by $H_{out}$ ('outside'). We also label oxygen atoms in the first solvation shell 'inside' and those in the second solvation shell 'outside.' We define the coordination number of the electron, $n_H$, as the number of $H_{in}$ protons. Histograms of the quantities $n_H$ and $n$ are shown in diagrams (i) and (ii) of Figure 1a, respectively. Based on the snapshots chosen from our one-electron MQC MD simulations, we find that the cavity electron is coordinated by 3-8 hydroxyl groups (with a mean coordination number $\langle n_H \rangle$ of ca. 6) inside a cluster of 12 to 25 water molecules (with an $\langle n \rangle$ of ca. 20 molecules).

To better characterize the snapshots chosen for this study, in Figure 1S(a) we show the distribution function $g(r_{XH})$ for the $r_{XH}$ distances, which has its first maximum at 2.26 Å. This pair distribution function is similar to the one obtained in identical MQC MD simulations by Schwartz and Rossky using a much larger sample of $e_{hyd}^-$ configurations. [9] In Figure 1S(b), we plot a histogram for the smallest of the $X$-$O$-$H_{in}$ angles for water molecules in the first solvation shell. The most probable value of this angle is 12-14° and the largest such angle is still less than 60°: the OH bonds in the first solvation shell clearly point preferentially towards $X$, consistent with previous conclusions from the literature. [8,9] Figure 2 displays the $X$-$H$ and $X$-$O$ histograms for 'inside' and 'outside' atoms, as defined above, and Table 1 summarizes the mean values of the $X$-$H_{in}$ and $X$-$H_{out}$ distances, which are 2.4 Å and 3.4 Å, respectively. We note that the mean $X$-$H_{in}$ distance from these one-electron snapshots is considerably longer than the 2.1-2.2 Å distance suggested by magnetic resonance experiments on alkaline ices. [2,4,6]

### 2.2. DFT and CIS calculations.

The electronic structure of the embedded water cluster anion snapshots described in the previous section was first modeled using unrestricted DFT with the B3LYP functional (Becke's exchange functional [27] and the correlation functional of Lee, Yang, and Parr) [28] from Gaussian 98 and 03, as noted. [29] The justification for using this implementation of DFT as opposed to *ab initio* methods for calculating magnetic resonance information was provided in Part 1 of this study. [1] Unless otherwise specified, for all of our DFT calculations, a 6-31G split-valence double-$\zeta$ Gaussian basis set augmented with diffuse and polarized (d,p) functions (6-311++G**) was used, [29] with a ghost 'chlorine' atom placed at the electron's center of mass, $X$. These calculations (using the standard *Prop=EPR* routine in Gaussian 03) [29] yielded *isotropic* hfcc's $a^{H,O}$ for the [1]H nuclei (the hfcc's on deuterons are 6.5 times smaller) and [17]O nuclei, defined as [30]

$$a = \left( \frac{8\pi}{3} \right) g_e \beta_e g_n \beta_n \rho(0), \tag{1}$$

where $g_e$ and $g_n$ are the electron and the nuclear g-factors and $\beta_e$ and $\beta_n$ are the corresponding magnetons and $\rho(0)$ is the spin density on the nucleus, and also provided the *anisotropic* hyperfine coupling tensors **B** (that is, the electron-nucleus magnetic dipole interaction) defined through [30]

$$\mathbf{B}_{ik} = g_e \beta_e g_n \beta_N \left\langle r^{-5} \left( 3 r_i r_k - r^2 \delta_{ik} \right) \right\rangle \; , \tag{2}$$

where $r_i$ is the Cartesian component of the radius vector $\mathbf{r}$ pointing from the nucleus to the electron and $\langle \; \rangle$ stands for averaging over the unpaired electron density. These traceless hyperfine tensors, with principal values of $(B_{xx}, B_{yy}, B_{zz})$, were nearly axial, so that $B_{xx} \approx B_{yy} = T_{\perp}$ and $B_{zz} = -2T_{\perp}$ [6] (observe that for $^1$H and $^2$H, $B_{zz}^H > 0$, whereas for $^{17}$O, $B_{zz}^O < 0$ because the nuclear moment for $^{17}$O is negative). Below, the hfcc are given in units of Gauss (1 G = $10^{-4}$ T); to convert these constants to frequency units (MHz), they should be multiplied by 2.8. These hfcc data also were used to calculate (using the method detailed in Appendix A in ref. 1) (i) the second moments ($M_2^{O,H}$) of the EPR spectra from the $^1$H and $^{17}$O nuclei, respectively, (eq. (A7) therein), (ii) the EPR spectra themselves (eq. (A3) therein), and (iii) ESEEM spectra (eqs. (A12) to (A16) therein). We also used Mulliken population analysis to determine the atomic spin ($\rho_s^{H,O}$) and charge ($\rho_c^{H,O}$) density on the corresponding atoms; all of the calculated parameters from our hybrid DFT-MQC MD model calculations are given in Table 1.

The IR and Raman spectra of embedded clusters and individual water molecules were calculated for 400 snapshots using the DFT/6-31+G** method and the standard *Freq=Raman* routine in Gaussian 98. [29] It is important to note that since the water molecules in the embedded clusters are not at their stationary points, the frequencies calculated from diagonalization of the Hessian matrix correspond to making a local harmonic, or instantaneous normal mode, approximation. Although our use of such an approximation may decrease the fidelity of our vibrational analysis, we know of no obvious way in which this shortcoming of our hybrid model can be overcome. Once we completed the locally harmonic analysis, the resulting 'line' spectra (for normal modes only) were binned (with the bin width set to 50 cm$^{-1}$) to produce the spectra shown in section 4.3. These line spectra were used to calculate centroids $\langle \nu \rangle$ of a given band $(\nu_{\min}, \nu_{\max})$ (see Table 2) defined as

$$\langle \nu \rangle = \int_{\nu_{\min}}^{\nu_{\max}} d\nu \; \nu \; I(\nu) \Bigg/ \int_{\nu_{\min}}^{\nu_{\max}} d\nu \; I(\nu) \tag{3}$$

where $I(\nu)$ is the calculated intensity (this binning is illustrated in Figure 15S in section 3.4). Note that the centroids calculated for the IR and Raman bands are different (Table 2). Although our ensemble of snapshots is too small to obtain high-quality IR-Raman spectra, it was sufficient to locate the band centroids with the accuracy of several cm$^{-1}$, as we found by comparing the centroids calculated using eq. (3) with centroids from smaller subensembles.

When comparing our calculated vibrational band centroids to experiment, it is important to note that the experimental data are *resonance* Raman spectra, [19,20] whereas our simulated spectra are regular IR and Raman spectra. Consequently, for our large embedded water anion clusters, both the vibrations of the OH groups forming the cavity and the vibrations in water molecules in the second solvation shell are present in the spectrum, whereas the experimental resonance Raman spectrum selects only those modes that are significantly displaced upon electronic excitation of the electron, which are presumably those of only the first-shell water molecules. Since the calculation of a *resonance* Raman spectrum in our hybrid method was not feasible with the available computational resources, we chose to examine only relatively small clusters with $r_{cut} = 3$ Å in order to selectively observe the vibrations of water molecules in the first solvation shell. We also note that the process of embedding an explicitly treated cluster into a matrix of point charges changes the calculated vibration frequencies because the water molecules at the surface of the cluster interact with these point charges.

For simulation of electronic spectra, our CIS calculations included either the first 10 or 20 excited states, denoted as CIS(N=10) and CIS(N=20) respectively, which were used to calculate both transition dipole moments and transition energies for the absorption spectrum of $e_{hyd}^{-}$ (in these CIS calculations, either a reduced 6-31+G* or 6-31+G** basis set was used, including an optional ghost 'Cl' atom at the electron's center of mass). Oscillator strengths were calculated by averaging the line spectrum from each configuration over the ensemble of snapshots as well as binning the transitions in frequency space using a bin size of 0.1-0.2 eV.

### 3. Results.

### *3.1. The SOMO (HOMO).*

An important part of our DFT analysis was the examination of the orbital structure, and in particular, the singly-occupied molecular orbital (SOMO), $\Psi$, which is also the HOMO. In one-electron models, the wavefunction of the *s*-like ground state of the $e^-_{hyd}$ is contained almost entirely within the solvation cavity. [8-12] In our multielectron hybrid DFT model, however, we observe that the SOMO is shared between the cavity and the *O 2p* orbitals of the first-shell water molecules. Typical isodensity surface maps of the SOMO for two sequential snapshots are shown in Figure 3, and a larger sample of such maps is given in the Supplement, Part B. Examination of Figure 3 (and the model clusters examined in ref. 1) indicates that in addition to the cavity, part of the SOMO occupies the frontal lobes of *2p* orbitals of the oxygen atoms in the first solvation shell. Moreover, the wavefunction inside the cavity and in these frontal lobes have opposite signs, so in the following we choose a phase convention so that the intracavity SOMO is positive. In this regard, the SOMOs from Gaussian 98 shown in Figure 3 are consistent with previous *ab initio* and DFT calculations for gas phase water anions that internally trap an electron. [1,15,16] On the other hand, the recent CPMD simulations of the $e^-_{hyd}$ by Boero et al. [17] showed no such features. This CPMD calculation also located the first peak of the $g(r_{XH})$ distribution for the protons at 1.6 Å, [17] which is significantly shorter than 2.0-to-2.3 Å obtained in MQC MD, [9] path integral, [8] and mobile Gaussian orbital [11] molecular dynamics calculations. We believe that the unusual structure of the $e^-_{hyd}$ in Boero et al.'s CPMD calculations may stem from the choice of pseudopotentials used in these calculations, the small box size (only 32 water molecules), and the method used for charge screening.

Using the fact that most of the SOMO density on the water molecules is contained in the frontal lobes of *O 2p* orbital and has opposite phase to the SOMO density in the cavity, we found it useful to define "positive" and "negative" charge densities via

$$\rho_{\pm} = \int d^3\mathbf{r}\, \Psi(\mathbf{r})^2\, \theta(\pm\Psi)\, , \qquad\qquad (4)$$

where $\theta(\ )$ is the Heaviside step function. From the normalization condition of $\Psi$, it follows directly that $\rho_+ + \rho_- = 1$; our convention of choosing the positive phase inside the cavity gives $\rho_+ > \rho_-$. Figure 4a shows the histogram of $\rho_-$ over 200 snapshots: the negative part accounts for 10-14% of the total SOMO density with an expectation value of 12% (there is, of course, additional electron density in positive lobes of the O *2p* functions). Consistent with the conventional way in which the spin density in the *p*-orbitals is determined from experimental EPR data, [1,30] we estimated the total spin density, $\phi_{2p}^O$, in the *O 2p* orbitals of (several) water molecules from the sum $\sum_O B_{zz}^O / B_{zz}^O (at.)$ taken over all $^{17}O$ nuclei, where $B_{zz}^O (at.) \approx$ -104 G is the corresponding atomic constant (see also Part 1). [30] The advantage of quantifying the orbital overlap this way (as compared to, e.g., orbital decomposition into atomic wavefunctions) is that the tensor given by eq. (2) "filters out" the components of the correct symmetry and thus provides a local measure of the *p*-character. This calculation indicates an 18±2% total overlap of the SOMO with the O *2p* orbitals (see Figure 1b for the histogram of this quantity). Thus, the penetration of the hydrated electron's wavefunction into the water molecules of the first solvation shell is not negligible.

To better characterize the SOMO, we found it convenient to introduce the angle-averaged (radial) density $\rho(r)$ of the electron wavefunction, defined through equation

$$4\pi r^2 \rho(r) = \left\langle \int d\Omega\, r^2 \Psi^2(\mathbf{r}) \right\rangle \qquad (5)$$

where $\Omega$ represents the solid angle and the angled brackets indicate an average over the ensemble of snapshots. We plot the quantity $4\pi r^2 \rho(r)$ and its running integral over $r$ in Figure 4b, which shows that the most probable position of the excess electron is $r \approx 1.75$ Å, well inside the cavity given that the most probable *X-H*$_{in}$ distance is 2.26 Å. Figure 4b also shows that 50-60% of the spin density is contained within a 2.2-2.4 Å sphere and that 75% is contained within the 3 Å cutoff radius that we used to define the first shell of water molecules. The figure also shows the diffuseness of the excess electron's wavefunction: ca. 5% of the spin density is contained beyond the most probable location of the *H*$_{out}$ protons, most of which resides in the *2p* orbitals of oxygen atoms in the second solvation shell.

Despite the pronounced features between 2 and 3 Å (at which the lobes of the *O 2p* orbitals show up), the general outlook of the SOMO generated from the many-electron calculation is similar to that given by one-electron models.[12,13] To demonstrate this, we note that in the simplest semicontinuum models,[12] the ground-state *s*-function of $e_{hyd}^-$ is given by

$$\Psi_s(r) \propto \exp[-r/\lambda], \tag{6}$$

where $\lambda$ is the localization radius of the electron. Fitting the radial density shown in Figure 4b to $4\pi r^2 \Psi_s^2(r)$ gives the optimum $\lambda \approx 1.67$ Å, which is indeed close to the most probable location of the SOMO. As seen from Figure 3, the SOMO for each particular snapshot is highly irregular. To better characterize the general shape of the SOMO, we elected to use multipole analysis. At the $l = 2$ pole, we characterize the charge distribution by a symmetrical gyration tensor

$$\mathbf{G}_{ij} = \left\langle \mathbf{x}_i \mathbf{x}_j \right\rangle_\Psi - \left\langle \mathbf{x}_i \right\rangle_\Psi \left\langle \mathbf{x}_j \right\rangle_\Psi, \tag{7}$$

where $\mathbf{x}_i = \{x, y, z\}$ and $\langle\ \rangle_\Psi$ stands for averaging over the SOMO density. This tensor is related to the (potentially experimentally observable) diamagnetic susceptibility tensor $\boldsymbol{\chi}$ of the excess electron via $\boldsymbol{\chi} = e^2/4m_e c^2 \{\mathbf{G} - tr(\mathbf{G})\mathbf{1}\}$. The gyration tensor $\mathbf{G}$ has eigenvalues $(r_a^2, r_b^2, r_c^2)$, arranged so that $r_a < r_b < r_c$, that give the semiaxes of the gyration ellipsoid. The radius of gyration is then defined as

$$r_g^2 = r_a^2 + r_b^2 + r_c^2 = \left\langle r^2 \right\rangle_\Psi - \left\langle r \right\rangle_\Psi^2. \tag{8}$$

The shape of the ellipsoid also can be characterized using the mean meridianal ($e_m$) and polar ($e_p$) eccentricities, defined as

$$e_m^2 = 1 - r_a r_b / r_c^2 \text{ and } e_p^2 = 1 - r_a^2 / r_b^2, \tag{9}$$

where for a truly spherically-symmetric hydrogenic *s*-like wavefunction, the three semiaxes would be equal to the localization radius $\lambda$ so that $e_m = e_p = 0$. For the same ensemble of 200 snapshots, the mean radius of gyration $r_g \approx 2.75$ Å (vs. 2.04 Å in the MQC MD model) and the mean gyration ellipsoid is 1.48 Å x 1.58 Å x 1.69 Å (vs. 1.07 Å x 1.17 Å x 1.28 Å in the MQC

MD model).  Thus, the shortest and the largest semiaxes of the gyration ellipsoid **G**, in the DFT/MQC MD model, are within 7% of the mean value; the mean eccentricities are $e_m \approx 0.42$ and $e_p \approx 0.33$.  Distribution functions for the parameters of the gyration ellipsoid are shown in Figure 5; we see that the gyration radius varies between 2.5 and 3 Å.  We note that the mean values given above do not convey the degree of variation in the shape of the SOMO between snapshots: the principal semiaxes cover a wide range from 1.3 Å to 1.9 Å, and the eccentricities vary from 0.1 to 0.6, as shown in Figures 5b and 5c, respectively.  The correlation plot in Figure 2S(a) suggests that the radius of gyration scales roughly linearly with the mean $X$-$H_{in}$ distance, following $r_g \approx 1.13 \langle r_{XH} \rangle_{in}$.

The 2.75 Å average radius of gyration that we calculate is significantly greater than the experimental value of 2.5-2.6 Å estimated from moment analysis of the optical spectrum via eq. (10) [31]

$$r_g^2 \approx \frac{3\hbar^2}{2m_e} \int dE \, E^{-1} f(E) \Big/ \int dE \, f(E),$$  (10)

where $E = \hbar\omega$ is the transition energy and $f(E)$ is the oscillator strength of the corresponding electronic transition (see section 3.2); it also greatly exceeds the estimate of ca. 2.04 Å obtained directly from the MQC MD model using the Schnitker-Rossky electron-water pseudopotential. [30] Using the experimental estimate for $r_g$ and the correlation plot given in Figure 2S, one obtains a mean $X$-$H_{in}$ distance closer to 2.2 Å rather than the 2.4 Å given by our MQC MD model.  The EPR and ESEEM data also suggest smaller cavities than we calculate here (see ref. 1 and section 3.3).

### 3.2. Energetics and the absorption spectrum.

In one-electron models, the ellipticity of the solvation cavity has important consequences for the absorption spectrum in the visible: the presence of asphericity splits the triply-degenerate $p$-like excited states, leading to three overlapping subbands with orthogonal transition moments. [9] This important feature (like the overall particle-in-a-box character of the electron wavefunctions) is also found in both our hybrid DFT-MQC and CIS-MQC MD models. To see this, in this subsection, we begin by exploring the density of states (DOS) function. To obtain the DOS from the DFT calculations, we calculated histograms of the Kohn-Sham eigenvalues for the occupied and virtual eigenstates of both spin orientations (in our convention, the SOMO is an $\alpha$ function).

Figure 3S(a) shows the DOS computed this way for our embedded water cluster anions (with $r_{cut}$ = 4.75 Å). The DOS exhibits two features near the bottom of the 'conduction band' that are shown separately in Figure 6a. Feature (i) results from the highest occupied $\alpha$–orbital (the Kohn-Sham HOMO, which is also the SOMO) that is located ca. -1.69 eV below the vacuum energy (the DOS maximum is at -1.8 eV vs. -1.75 eV given by the CPMD calculation of Boero et al). [17] Feature (ii) derives from the three lowest unoccupied molecular $\alpha$–orbitals (LUMO, LUMO+1, and LUMO+2), which have centroids at 0.42, 0.65, and 0.86 eV, respectively. It is natural to make a correspondence between these three states and the three nondegenerate $p$-like states observed in one-electron models. [8-11] Given the correspondence between the multielectron DOS and the energetics observed in one-electron models, we expect that transitions from the SOMO to these three states will dominate the optical spectrum in the visible. To examine this, in Figure 6b, we plot histograms of the corresponding transition energies, which indeed show three distinctive $p$-subbands with centroids at 2.11, 2.34, and 2.55 eV; we note that these histograms are *not* identical with the spectra because we have not weighted these by their corresponding oscillator strengths. For comparison, path integral calculations by Schnitker et al. [8b] using the same pseudopotential as for our MQC MD calculations gave peak positions at 2.1, 2.5, and 2.9 eV. The path integral calculations yield an absorption spectrum that is shifted to the blue by 0.34 eV relative to the experimental one (shown by the dashed curve in Figure 7b), which is centered at $E_m \approx 1.7$ eV. [32] Note that energy of the HOMO generally increases (Figure 2S(b)) and the corresponding energy gaps between the HOMO and the lower unoccupied states decrease (Figure 2S(c)) with the increasing radius of gyration. That is, the spread in these energies is largely accounted for by the variation in the cavity size. The same anticorrelation was observed in the MQC calculations of Coudert and Boutin, [33] for $e_{hyd}^-$ in nanoconfined water pools in zeolites, and Rossky and Schnitker, [23] for the hydrated electron in bulk water.

Figures 7a and 7b show the absorption spectrum calculated using CIS($N$=10)/6-31+G* method (section 2) for our embedded water anion clusters with $r_{cut}$ =4.75 Å. We note that the excited states of $e^-_{hyd}$ are substantially more diffuse than the ground state (see the discussion below). Thus, to ensure that these CIS calculations are reliable for these excited states, more than two water shells need to be treated explicitly with our embedding method. Unfortunately, including additional water shells is not feasible because of the excessive computational demands of such a large calculation. Even though we are not certain that the excited-state wavefunctions are converged when only two solvent shells are treated explicitly, we can nevertheless draw some conclusions from these calculations, as the spectra of $e^-_{hyd}$ calculated via CIS using a ghost atom essentially does not change when either one and two surrounding water shells are explicitly included (Figure 4S(a)). Thus, for the largest CIS calculation we can perform, the calculated absorption spectrum appears not to be sensitive to the size of the embedded cluster.

Despite this lack of size sensitivity, the CIS-calculated spectrum is still significantly blue-shifted relative to experiment (Figure 7b). This blue shift is likely the result of the level of theory that we use (CIS($N$=10)/6-31+G*). Specifically, this CIS calculation does not correctly reproduce the electrostatics of bulk water. The typical Mulliken charge on the protons from our CIS calculations is $Q_H$=0.55, which is substantially greater than SPCf model charge of $Q_H$=0.41 that is known to reproduce the experimental dipole in the liquid and agrees well with the DFT calculations. To understand the effects of CIS generating too large a dipole for liquid water, we performed single-electron MQC calculations using an artificially large dipole moment for the surrounding water molecules and we found that increasing the water dipole to that seen in the CIS calculations increases the *s-p* gap of the hydrated electron by ~ 0.3 eV. In addition, we note that single-excitations in the CIS technique are not sufficient to allow the water molecules to polarize appropriately in the presence of the excited states of $e^-_{hyd}$, so the very fact that CIS includes only single excitations also tends to increase the s-p gap. Overall, the best we can infer from the CIS spectra shown in Figure 7 is that it can reproduce the visible absorption spectrum of the hydrated electron about as well as traditional one-electron MQC calculations [8,23] that also exhibit significant blue shift (see above). A spectrum that more closely resembles the experimental one (including the characteristic "tail" in the blue) was obtained in a CIS($N$=20)/6-31+G** calculation that included only one complete solvation shell of water molecules and no ghost atom (Figure 5S(a)). Furthermore, almost perfect agreement with the experiment was obtained when in the latter calculation the matrix of point charges was removed (Figure 5S(b)). This illustrates the great sensitivity of the calculated CIS spectra to the details of cluster embedding and the choice of the basis set. This sensitivity, in turn, is explained by the large spatial extent of the excited states and the difficulty in representing the outer parts of the electron's excited-state wavefunctions correctly.

With the above caveats in mind, the calculated CIS spectra, despite considerable sharing of the excess electron into the *O 2p* orbitals, still exhibit the features that are observed in (one-electron) path integral and MQC MD calculations. The three subbands centered at 2.09, 2.43, and 2.76 eV (Figure 7a) correspond to the three lowest excited states that have nearly orthogonal transition dipole moments (see the inset in Figure 7b). Each one of these subbands carries an integral oscillator strength of ca. 0.3. The total integrated oscillator strength is ca. 0.95; for the CIS($N$=20)/6-31+G** calculation with a single solvent shell, shown in Figure 5S(b), it is ca. 1.15. There is an anticorrelation between the transition dipole moment and the transition energy (Figure 4S(b)) that is also seen in mobile Gaussian orbital set calculations of Borgis and Staib. [11] The estimates of $r_g$ for the radius of gyration obtained using eq. (10) are 2.14 Å (Figure 7b) and 2.25 Å (Figure 5S(a)) (both of which are considerably lower than the direct estimate for this parameter obtained using eq. (8) (see section 3.1).

We can also compare the excited states from the CIS calculations with the Kohn-Sham orbitals obtained in the DFT calculations. In Figure 6S, we show isodensity contour plots from our DFT calculations of the Kohn-Sham HOMO, HOMO-1, LUMO, LUMO+1, and LUMO+2 for one of the snapshots; we find that all such isodensity surfaces are qualitatively similar. A sequence of such plots for the LUMO (from a different snapshot than that in Figure 6S) as a function of density level is shown in Figure 7S. The familiar dumbbell shape of the '*p*-orbital' is not readily recognizable, although the three lower unoccupied states do exhibit *p*-like polarization, each orthogonal to the others (Figure 6S). Nevertheless the orbital structure of these '*p*-states' is rather different from that obtained in MQC MD models [9] (and the CPMD model): [17] only a fraction of the total '*p*-state' density (ca. 20%, Figure 8S(a)) is contained inside the cavity. The *p*-character of these electronic states is achieved mainly through the polarization of the frontal *O 2p* orbitals in the OH groups forming the cavity: the phase of the electron in these orbitals on one side of the cavity assumes a positive sign, while the phase of the electron in the *O 2p* orbitals straight across the cavity in the direction of the transition dipole moment assumes a negative sign. In addition, we also see both positive and negative excess electron density in the interstitial cavities between the water molecules of the first and the second solvation shells. Thus, in our multi-electron calculations, we see that the excited '*p*-like states' of the $e_{hyd}^-$ extend further out of the cavity than the '*s*-like' ground state. To illustrate this, angle-averaged densities for the LUMO, LUMO+1, and LUMO+2 are plotted in Figure 8S(a). The corresponding gyration ellipsoid for these orbitals is 1.8 Å x 2.2 Å x 3.3 Å (the distribution of semiaxes is shown in Figure 8S(b)), making them nearly twice the size of the gyration ellipsoid for the SOMO (section 3.1). The mean meridianal eccentricity of these three excited states is ca. 0.79, which is close to the theoretical 0.75 for a *p*-orbital, and the mean radius of gyration is $r_g \approx$ 4.33 Å (vs. 2.75 Å for the HOMO). Once again, despite the complex orbital structure and the crucial involvement of *O 2p* orbitals, we observe that on average, the lowest unoccupied molecular orbitals in the DFT calculations still resemble the *p*-like states given by one-electron models.

By contrast, a novel feature of our DFT calculations that is *not* captured by one-electron models can be seen in Figure 3S(a), which shows an onset (feature (iii)) in the density of states that results from the occupied *O 2p* orbitals in the water molecules. This onset arises from a band of HOMO-1 orbitals that are composed of *1b₁* orbitals of the water molecules in the first solvation shell; a typical such HOMO-1 orbital is shown in Figure 5S(b). Our calculations suggest that the onset of this band starts 5.75 eV below the vacuum level and it has its first peak at –7.5 eV. The presence of this peak suggests that there should be an electronic transition from the occupied *O 2p* orbitals into the HOMO at ca. 5.95 eV (~210 nm). In fact, the experimentally observed UV band of the hydrated electron peaks at 6.5 eV (190 nm) with an onset around 220 nm. [18] Thus, our hybrid DFT-MQC MD model provides an assignment for the observed UV band of the hydrated electron. [We note that our assignment of the band as involving transitions from the *O 2p* orbitals of water molecules in the first solvation shell was also given speculatively by Hart and co-workers in 1976]. [18] The CIS method cannot reproduce this 190 nm band because the limited size and number of excited states in our calculation excludes the possibility of excitations of the core electrons on the adjacent water molecules.

To better understand the origin of these transitions, in Figure 3S(b) we plot both the DOS function for 200 embedded *neutral* clusters (using $r_{cut} = 3.5$ Å from a central oxygen atom) and the DOS for the anionic clusters including only a point negative charge at $X$ (instead of a full description of the excess electron); we also include the DOS of embedded small water anions using only the first solvation shell ($r_{cut} = 3$ Å). Figure 3S shows that the DOS of both the neutral and small anion clusters have three peaks at –13 eV, -9.8 eV, and –7.6 eV that correspond to the *1b₂*, *3a₁*, and *1b₁* orbitals of *neutral* water molecules (see Figure 20 in ref. 40 for a sketch of these orbitals); the respective bands for these orbitals have been observed experimentally in the photoelectron spectra of liquid water by Faubel and co-workers. [34] Figure 3S also shows that the presence of a point negative charge in the cavity causes a Stark shift of the eigenvalues towards the midgap by ca. 1.1 eV. The most prominent feature in the DOS of the anionic clusters corresponds to the upshifted *1b₁* band (that is, the HOMO-1 orbital) that arises from the *O 2p* orbital in the water molecule that is perpendicular to its plane (Figure 6S(b)). It turns out that even a point charge placed at the cavity center can fully account for this upshift; a full wavefunction description of the intracavity electron does not significantly change the calculated DOS for occupied states with energies more negative than -3 eV.

To summarize this section, our hybrid DFT- and CIS-MQC MD calculations qualitatively account for many of the experimentally observed features of the hydrated electron, including its absorption bands in the visible and UV. The three lower unoccupied states are nondegenerate and correspond to *p*-type orbitals oriented along the three principal axes of the electron's elliptical cavity (section 3.1). The splitting between the corresponding *p*-subbands and their widths are comparable to the those reported in both MQC MD [9] and CPMD calculations [17] despite the qualitative differences between these three models. Although the DFT calculations yield rather different orbital structure for the '*p*-states' (Figures 6S and 7S) than the MQC MD and other one-electron models, [8-11] the absorption spectrum still resembles the one simulated using such one-electron models. Thus, our results suggest that the optical spectrum alone cannot be used to validate or invalidate models for the $e^-_{hyd}$. Instead, other experimental features, such as hfcc parameters determined using magnetic resonance methods, [1-6] or vibrational parameters obtained from resonance Raman [19,20] are needed to refine our theoretical understanding of the hydrated electron.

### 3.3. EPR and ESEEM spectra

EPR and ESEEM spectroscopy provide estimates for hfcc's that strongly depend on the cavity geometry and amount of spin density of the excess electron overlapping with magnetic nuclei ($^1H$, $^2H$, and $^{17}O$) in the nearby water molecules. In Part 1 of this study, [1] we found that in water cluster anions that trap the electron internally, the excess spin and charge density are localized mainly on the OH groups of the first solvation shell. For the embedded cluster anions examined in the present study, this same type of distribution is also seen, as documented in Figure 9S, which exhibits histograms of the atomic spin ($\rho_s^{H,O}$, panel (a)) and charge ($\rho_c^{H,O}$, panel (b)) densities for our $r_{cut} = 4.75$ Å clusters. For comparison, Figure 9S(b) shows the charge distribution on individual neutral water molecules in the matrix (with $\langle \rho_c^H \rangle \approx 0.36$). Examination of this latter plot suggests that for the anionic water clusters, both the O atoms in the second solvation shell and the $H_{out}$ atoms in first solvation shell have an atomic charge that is within 0.02 $e$ of what is observed for bulk water molecules. For the solvating OH groups, however, the charge on the $H_{in}$ hydrogens is 0.2 more negative than in neutral water, and the charge on the first-shell oxygen atoms is 0.17 more positive than in neutral water. Figure 9S(a) also shows that the spin density follows a similar trend: the spin density on the $H_{out}$ nuclei is small and on the oxygen nuclei in the second shell is almost negligible; the most probable values for the spin density on the $H_{in}$ and the first-shell O atoms are +0.1±0.05 and -0.04±0.01, respectively. It is noteworthy that the distribution of the spin density for the $H_{in}$ atoms is very broad, spanning a range from -0.2 to +0.4.

Figures 8a and 8b show histograms (1000 snapshot average) of the isotropic and anisotropic components of the hyperfine coupling tensor for $H_{in}$ and $H_{out}$ protons and $^{17}O$ nuclei, respectively (see section 2.2 for the definition of classes of the nuclei). This figure illustrates the difficulty of finding a 'representative' water cluster that describes the hydrated electron (such as, for example, "Kevan's" octahedral model examined in Part 1): [1] there is a broad distribution of calculated hfcc values, and the distributions for the isotropic hfcc's are skewed, so the mean values are quite different from the most probable ones (Table 1). The mean values of $a^O$ for oxygen-17 nuclei in the first and the second solvation shells are ca. -15 G and ca. -2.1 G, respectively. The correlation plot of these isotropic hfcc's vs. X-O distance, given in Figure 10S(a), shows that to a first approximation $a^O \propto \exp(-2\,r_{XO}/\lambda_O)$, where $\lambda_O \approx 1.59$ Å is close to the localization radius $\lambda$ of the SOMO (see section 3.1 and eq. (6)). No such correlation is obvious for the protons since the hfcc depends on the orientation of the $O\,2p$ orbital to which the electron in the $H\,1s$ orbital is coupled. The correlation plot of $B_{zz}^H$ for the $H_{in}$ protons given in Figure 10S(b) shows that the point dipole approximation, $B_{zz}^H (G) \approx 57.6/r_{XH}^3$ [1] (for X-$H_{in}$ distances in units of Å), holds well for $B_{zz}^H < 6$ G. This is due to both the relative sphericity of the SOMO (with its large mean coordination number $n_H$ ) and the preferential orientation of OH groups towards the cavity; by contrast, in model anion clusters with low coordination number of the cavity electron and $d$-orientation of water molecules (see Part 1), [1] there is considerable deviation from the point dipole approximation.

Using the calculated hfc tensors, one can estimate the contributions from both $^1$H and $^{17}$O nuclei to the second moment, $M_2^{H,O}$, of the EPR spectra (Figure 11S). The contribution to this moment from each magnetic nucleus is given by $\frac{1}{3} I(I+1)\left(a^2 + 2T_\perp^2\right)$, where $I$ is the nuclear spin. For protons (deuterons) the second, dipolar, term in the last factor prevails. For NaOH:$H_2$O glasses, the proton contribution, $M_2^H$, was determined experimentally to be between 21 and 23 G$^2$. [4,6] Our hybrid calculation gives a mean value of 17.3 G$^2$ (a histogram of our calculated values for $M_2^H$ is shown in Figure 11S(b)); ca. 80% of the mean value comes from the anisotropic hyperfine interaction. There is also an additional contribution to $M_2^H$ of ca. 0.8 G from the remote matrix protons, which can be treated using the point-dipole approximation. Our slightly low estimate for $M_2^H$ is likely due to the overestimated cavity size from our MQC MD simulations, which gives $\langle r_{XH} \rangle_{in}$ = 2.4 Å for the $H_{in}$ protons; in contrast, experimental estimates for the cavity size from EPR [4] and ESEEM [2,6] are 2-2.2 Å. The fact that our calculations are based on clusters with too large a cavity also results in a relatively low estimate for $\langle B_{zz}^H \rangle_{in} \approx 3.7$ G; the experimental estimates of this quantity are 6-7 G. [2,6] It is worth noting that due to the very steep (cubic) dependence of the dipole component of the hfcc tensor on $r_{XH}$, even a small error in the cavity size causes a large error in the estimates for $B_{zz}^H$. (The fact that our cavity size is likely too large is also evident when comparing our calculated radius of gyration $r_g$ for the electron to experimental estimates, as noted in section 3.1).

Figure 12S(a) shows the simulated EPR spectrum, which is close to Gaussian in shape and which looks much like the experimental spectrum in the alkaline glasses.[4] The peak-to-peak line width for $\Delta B_{pp}$ of 9.1 G that we calculate compares well with the experimental estimate of 9.5±0.5 G reported by Astashkin et al.[6]

For $^{17}$O nuclei, the second moment of the EPR spectrum is ca. $6 \times 10^3$ G$^2$, almost all of which is due to the *isotropic* hyperfine interaction (see above). For the 37% oxygen-17 enriched sample studied by Schlick et al.,[5a] using eq. (A7) in ref. 1 we obtained $M_2 \approx 2250$ G$^2$ vs. the reported experimental estimate of 134 G$^2$. As we observed in Part 1,[1] *all* ab initio and DFT models of the hydrated electron tend to give such large estimates for $M_2^O$. We note, however, that the experimental estimate of Schlick et al.[5a] was compromised by their subsequent observation[5b] of a strong spectral overlap between one of the resonance lines of the O$^-$ radical and the narrow EPR signal from the "electron," which had $\Delta B_{pp} \approx 18 \pm 1$ G. In alkaline glasses, the O$^-$ anion is formed with the same yield in the same radiolytic reaction that yields the $e_{hyd}^-$. In $^{16}$O glasses, the two narrow EPR signals from $e_{hyd}^-$ and O$^-$ are spectrally well separated, but because the signals overlap in $^{17}$O enriched samples, the EPR spectrum in such enriched samples is very complex: there are 7 lines from $^{17}$O$^-$ spanning 400 G with the $g_{II}$ component (Figure 1 in ref. 5b) strongly overlapping with the EPR signal from the electron. Thus, the small $M_2$ estimate of 134 G$^2$ given by Schlick et al., that was obtained using $^{17}$O enriched samples, is subject to some doubt.

To better understand the EPR spectrum of the enriched samples, in Figure 12S(b), we used our calculated hfc tensors for $^1$H and $^{17}$O nuclei to simulate the EPR spectrum of an oxygen-17 enriched sample. We find that the EPR line decomposes into two distinct spectral contributions, a narrow one with $\Delta B_{pp} \approx 23$ G and $M_2 \approx 135$ G$^2$ (in good agreement with the estimates of Schlick et al.[5a]) and a very broad line with $\Delta B_{pp} \approx 89$ G and $M_2 \approx 1980$ G$^2$. For a sample with 37% $^{17}$O enrichment, there is a ca. 10% probability that the first solvation shell would have no magnetic oxygen-17 nuclei. We assign the narrow line as arising from such isotopic configurations, so that the electron is only weakly coupled to the oxygen-17 nuclei in the second solvation shell. The isotope configurations that include at least one oxygen-17 nucleus in the first solvation shell, on the other hand, are responsible for the broad line. It is worth noting that our simulation in Figure 12S(b) neglects any differences in the paramagnetic relaxation of these two kinds of hydrated electrons. Small-amplitude movement of water molecules in the frozen samples would cause efficient spin relaxation, due both to the large hfcc's on the first-shell oxygens and the steep dependence of the isotropic hfcc on the X-O distance (Figure 10S(a)). The narrow line from the electron in the 37% $^{17}$O enriched sample is superimposed on a much broader signal that was attributed to one of the components of the $^{17}$O$^-$ radical multiplet (that shows a complex pattern of broad lines). This narrow resonance line was recognized as the EPR signal originating from the $e_{hyd}^-$ from the microwave power saturation behavior of the spin transition, i.e., this line was selected by its long relaxation time. Broad resonance lines that were not saturated were attributed to the $^{17}$O$^-$ radical. Such a criterion for the recognition of the (tentative) EPR signal from the $e_{hyd}^-$ discriminates against broad EPR signals with short relaxation times that are expected for trapped electrons that are strongly coupled to $^{17}$O in the first solvation shell. We believe, therefore, that the EPR results of Schlick et al. [5] do not contradict our hybrid DFT model, as the experimental results may be accounted for by assuming that only weakly coupled electrons (for which the magnetic $^{17}$O nuclei are in the second solvation shell only) are selected using the *ad hoc* criterion suggested by Schlick et al.: [5b] electrons that are strongly coupled to the $^{17}$O nuclei in the first solvation shell relax rapidly and have broad EPR lines that are superimposed on the comparably broad lines from the $^{17}$O$^-$ radical.

We turn next to the experimental ESEEM results for NaOD:D$_2$O glasses, which are shown in Figure 9 (see Appendix A of Part 1 for a brief introduction to ESEEM spectroscopy and how such spectra are simulated). Figure 9(a) shows simulated Fourier Transform modulo primary (p-) ESEEM spectra; in these calculations, we did not consider the (magnetically) weakly coupled matrix deuterons. These calculations reflect the "ideal" spectra that would be obtained assuming that the EPR spectrometer has no dead time and that there is no relaxation that narrows the observation window. In Figure 9(c), however, we include such effects, assuming a dead time of 250 ns and a relaxation time of 2 μs, as in the experiments of Astashkin et al.) [6] For comparison, the experimental spectra are shown in Figure 9(b). Since the weakly coupled matrix nuclei yield very strong signals at $\nu_D$ and $2\nu_D$ (where $\nu_D \approx 2.2$ MHz is the NMR frequency of the deuteron in the magnetic field of EPR spectrometer), Astashkin et al. [6] used rejection filtering at these frequencies to single out the more strongly coupled $D_{in}$ nuclei. These authors also used additional rejection filtering at a frequency of 7.6 MHz to reduce the (weak) modulation signal from magnetic $^{23}$Na nuclei; we note, however, that all the various filtering also serves to distort the spectrum. The two peaks marked (i) and (ii) in panel (b) of Figure 9 correspond to $v_{\alpha,\beta}^{\perp} \approx \left| v_D \pm \left( a^D + T_{\perp}^D \right)/2 \right|$ and the sum frequency $v_\alpha + v_\beta \approx 2v_D \left( 1 + \left[ T_{\perp}^D / 2v_D \right]^2 \right)$, where the indices $\alpha$ and $\beta$ refer to the two orientations of the electron spin with respect to the magnetic field. Since the parameters $a^D$ and $T_{\perp}^D \approx -B_{zz}^D/2$ are widely distributed (see Figure 8a for the proton hfcc's) the corresponding lines are broad. The reported estimates of "mean" hfcc's correspond to the positions of peaks that are not well defined for such broad lines. Simulation of the FT pESEEM spectrum using such "mean" values for one, two, [6] or six [2] magnetically equivalent deuterons do not yield traces that resemble the experimental ones (Figure A1 in ref. 1 and ref. 6).

By contrast, our simulation based on our hybrid many-electron calculation involves all of the deuterons coupled to the electron and averages over all such configurations. To distinguish the contributions from different groups of nuclei, Figures 9(a,c) show three traces plotted together: a trace for all deuterons in the first and the second solvation shells, a trace for only the $D_{in}$ nuclei, and a trace for just the close-in $D_{in}$ nuclei with $r_{XD} < 2$ Å (for which $\langle a^H \rangle \approx +0.68$ G and $\langle B_{zz}^H \rangle \approx 5.6$ G in the protonated sample). Since the deuterons in the second solvation shell are only weakly coupled, these nuclei give narrow lines in the FT pESEEM spectra that are superimposed on the broad signals from the $D_{in}$ nuclei. At high frequencies, there are also two lines (marked (iii) and (iv) that correspond to the third harmonic of the NMR frequency and a combination frequency, respectively, which are clearly seen in panel (b). These characteristic features were missing from all previous reconstructions of the ESEEM spectra. We note that no rejection filtering was used for our *simulated* ESEEM spectra, so the sharp lines from the distant deuterons at $v_D$ and $2v_D$ are superimposed on the broader lines of $D_{in}$ deuterons. As seen from the comparison of panels (b) and (c), the simulated FT pESEEM spectrum matches the experimental spectrum in the overall shape, although the center bands are shifted to lower frequency. This is, once again, a consequence of overestimated cavity size in the MQC MD model. Longer X-$D_a$ distances result in smaller dipolar coupling (Figure 6S(b)) and thus a lower $\nu_\alpha + \nu_\beta$ frequency. By choosing only those nuclei for which $r_{XD} < 2$ Å, however, it is possible to match the simulated and the experimental ESEEM spectra both in the positions of bands (i) and (ii) and their widths, as seen in Figure 9(c)). The results obtained in Part 1 of this study [1] for model octahedral clusters suggest that the matching is possible only for X-D$_{in}$ distances of 1.9-2.2 Å. Although the difference between this optimum distance of $r_{XH}$ and the MQC MD average of 2.4 Å is small, it is sufficient to reduce $T_\perp^D$, which steeply depends on this distance. Other than this, our hybrid DFT/MQC MD model appears to account for all of the experimentally observed features of EPR and ESEEM spectra of the $e_{hyd}^-$ trapped in alkaline ice.

### 3.4. Vibrational Spectra

In resonance Raman spectroscopy, only those vibrational modes that are significantly displaced upon electronic excitation show resonance enhancement; [19,20] thus, resonance Raman spectroscopy provides an excellent probe of the water molecules in the immediate vicinity of the hydrated electron. The vibrational peaks in the experimental resonance Raman spectra of the $e_{hyd}^-$ all exhibit significant downshifts relative to the peaks present in neat water without an excess electron. According to Tauber and Mathies, [19] the resonance Raman peak position for the $e_{hyd}^-$ in H$_2$O (vs. those for bulk water), in cm$^{-1}$, are: librations at 410 (vs. 425-450), 530 (vs. 530-590), 698 (vs. 715-766); the H-O-H bend at 1610 (vs. 1640); and the H-O stretches at 3100 (3420). Thus, the downshift of the bend mode, which exhibits a narrow, symmetric line, is ca. 30 cm$^{-1}$, and the downshift of the stretch mode (as estimated by the shift of the half-maximum of the broad, asymmetric line) is ca. 200 cm$^{-1}$. The question we address in this section is how to account for these experimentally observed vibrational downshifts.

There have been previous *ab initio* and DFT calculations for gas phase $(H_2O)_n^-$ anions [15,16] and related $(H_3O)(H_2O)_n$ [35] clusters in which the experimentally observed downshifts for the bending and stretching modes were qualitatively reproduced, suggesting that electron sharing with the nearby water molecules is the key to understanding the downshifts. One convenient way to demonstrate how sharing of the electron density by *O 2p* orbitals of water molecules affects the vibrations in these molecules is to examine the octahedral model of the hydrated electron (that is, a gas phase hexamer water anion that traps the electron internally; see Figure 1(a) in ref. 1). Thus, we elected to examine the vibrations of $^1H_2^{16}O$ molecules in the geometry-optimized hexamer anion clusters using DFT at the B3LYP/6-31+G** level; we note that with this level of DFT, the normal vibrational frequencies calculated for isolated water molecules are 1603 cm⁻¹ ($\nu_2$, H-O-H bend), 3800 cm⁻¹ ($\nu_1$, symmetric H-O stretch), and 3932 cm⁻¹ ($\nu_3$, asymmetric stretch), which agree only qualitatively with the experimental frequencies of 1595, 3657, and 3756 cm⁻¹, respectively.

In Figure 10, we plot the cavity size ($r_{XH}$) dependence of the vibrational band centroids of the water H-O-H bending and O-H stretching modes both for geometry-optimized octahedral anions, and for *neutral* octahedral clusters with the same $r_{XH}$. For the O-H stretch, we find that the downshift is linear with $\phi_{2p}^O$, defined in section 3.1 (Figure 13S), emphasizing that electron sharing with the first-shell water molecules is what is primarily responsible for the change in vibrational frequency. Figure 10 also shows that for $r_{XH} \approx$ 2-2.1 Å, the DFT calculations for the anion hexamers give ~20 and 200-250 cm⁻¹ downshifts for the band centroids of H-O-H bend and the O-H stretch, respectively; both of these estimates are in good agreement with experiment. [19]

Although they produce good agreement with experiment, the calculations summarized in Figure 10 are not entirely convincing because the hallmark of liquid water is its network of H-bonds, and this network is absent in our calculations for isolated gas-phase clusters. Thus, we examine the vibrational spectra of the $e_{hyd}^-$ calculated using our hybrid embedded cluster approach, even though such an approach (which is based on a local harmonic approximation, see section 2) is somewhat less rigorous than a full *ab initio* calculation. The vibrational spectra of the $e_{hyd}^-$ calculated using our hybrid embedded cluster approach are shown in Figures 11 and 12, and the vibrational band centroids are summarized in Table 2. Figure 11a shows the calculated IR spectrum for a 400-snapshot ensemble of water molecules generated in an MD simulation of neat SPCf [22] water (on which our MQC MD calculations are based). These embedded neutral water clusters show a narrow H-O-H bending band centered at 1747 cm⁻¹ and a broad O-H stretch band centered at 3127 cm⁻¹. We find that the exact positions of these two band centers depends sensitively on the assumed charge $Q_H$ on the matrix protons (Figure 13S(b)): the greater the charge, the lower the calculated frequency. For $Q_H$ =0.41 $e$ (the choice used in SPCf model), [22] the mean Mulliken atomic charge on the explicitly-treated protons is +0.36 $e$, leading to the vibrational frequencies summarized in Table 2. Traces (ii) and (iii) in Figure 11, respectively, show that even the vibrational bands of medium-size neutral clusters (whose size distribution is shown in Figure13S(a)) are somewhat different from those of embedded water molecules; it is clear that the vibrational frequencies we calculate for our embedded neat water model compare poorly with experiment.

Despite the fact that the absolute positions of the vibrational features calculated using our embedded water clusters do not match experiment, Figure 11 shows that the downshifts of these bands in the presence of the excess electron are well described by our hybrid calculation: we see downshifts of the librational modes at 750 cm$^{-1}$, the H-O-H bending mode at 1700-1750 cm$^{-1}$, and the O-H stretch modes around 3000 cm$^{-1}$ that are in reasonable agreement with the resonance Raman experiments. [19] To test whether the vibrational downshifts that we calculated originate simply through electrostatic interactions with the negative charge inside the cavity, we also calculated the IR-Raman spectra for embedded neutral clusters having exactly the same geometry as the water anion clusters with a point negative charge at $X$. Figure 12 presents a comparison of the vibrational spectra for these 'neutral' clusters (dashed curve) and 'point charge' clusters (solid curve); the corresponding band centroids are given in Table 2. The comparison clearly suggests that electrostatic interactions alone cannot account for the downshift of the vibration bands; it is the mixing of the excess electron's wavefunction with the frontier molecular orbitals of the first-shell water molecules that weakens the bonds and leads to the lower vibrational frequencies. The magnitude of our calculated downshifts can be estimated from the data in Table 2. The calculated downshift for the H-O-H bending mode is ca. 50-60 cm$^{-1}$ (as compared to the experimental estimate of 30 cm$^{-1}$); [19,20] the calculated downshift for the O-H stretching modes is 80-180 cm$^{-1}$ (as compared to the experimental estimate of 200 cm$^{-1}$ for the band center). [19] Thus, although our DFT estimates for band downshifts using *optimized* octahedral water clusters are closer to experiment, it appears that the hybrid embedded water anion model, despite its many approximations, accounts reasonably well for the observed features of the resonance Raman spectra of the $e^-_{hyd}$.

The plots of vibrational density of states (VDOS) for anion and neutral clusters (Figure 16S) reveal a large peak at 100 cm$^{-1}$ (3 THz) corresponding to low-amplitude motions of water molecules relative to the cavity (which is in the same 25-75 cm$^{-1}$ range as the symmetric "breathing" mode predicted by Copeland et al.[36]). No definitive signature of these low-frequency vibrations was found in autocorrelation functions for orbital and transition energies and the radius of gyration. Rapid exchange of water molecules occurring on the sub-picosecond time scale is the likely culprit.

## 4. Concluding remarks.

In this paper, we have shown that a combined DFT- and CIS-MQC MD approach can account, at least semiquantitatively, and in many instances, quantitatively, for many of the experimentally observed features of the ground-state $e_{hyd}^-$: these include the optical spectra in the visible and the UV, the vibrational spectra, and the EPR and ESEEM spectra. As far as we are aware, no other theoretical approaches suggested have provided such a comprehensive description of the hydrated electron's properties. The salient feature found from our multielectron model of the $e_{hyd}^-$ is that a considerable (ca. 18%) fraction of the excess electron's density resides in the frontal *O 2p* orbitals of oxygen atoms in the hydrating OH groups. Not only does this sharing not contradict any experimental observations, but this sharing naturally accounts for the magnetic resonance spectra and the downshifts of the water vibrational frequencies seen in resonance Raman experiments on the $e_{hyd}^-$.

Despite the extensive spin and charge sharing, the average ground and excited-state orbitals resemble in their general appearance the *s*- and *p*-like states in the one-electron models. This rationalizes the success of such models in explaining the energetics and main spectral features of the $e_{hyd}^-$. The DFT calculations also reproduce some other features seen in one-electron models, such as the facts that the gyration ellipsoid is aspherical, the lower three excited '*p*-states' are nondegenerate and extend beyond the solvation cavity, and that the transition dipole moments to these three '*p*-states' are nearly orthogonal. On the other hand, the nature of these '*p*-states' in our DFT calculations is somewhat different from that predicted by one-electron models; we find that the transition dipole moment results largely from the electronic polarization of the cavity (because half of the frontier *O 2p* orbitals assume positive and the other half negative sign). Our calculations also predict that photoexcitation changes the most probable electron distance from 1.72 Å from the cavity center to 3.3 Å, with most of the electron density contained in the interstitial voids between the water molecules of the first and the second solvation shells (only 20% of the excited electron's density resides inside the cavity, vs. the ca. 100% predicted from one-electron approaches). By contrast, 60% of the density of the ground '*s*-state' is confined inside the cavity and 80% is confined within the first solvation shell. This tight electron localization justifies the embedded cluster approach used in this study.

Our DFT calculations also provide an assignment for the 190 nm absorption band of the hydrated electron, [18] which has not been accounted for in one-electron models. Our calculations suggest that the presence of the negative charge inside the cavity makes the orbital energies of the valence electron in water molecules in the first solvation shell ca. 1.1 eV more positive than in liquid water. We thus assign the UV band of the hydrated electron as originating from an electron transition from a Stark-shifted *1b$_1$* orbital of the first-shell water molecules to the HOMO.

We also see that spin sharing of the excess electron by *O 2p* orbitals of the first and, to a lesser degree, the second solvation shells results in large hyperfine coupling constants for $^{17}$O nuclei in these molecules. We demonstrated how the seemingly contradictory EPR results of Schlick et al. [5] for $^{17}$O-enriched alkaline glass samples can be accounted for in the DFT model. It appears that fast paramagnetic relaxation and extreme broadening of EPR lines from the hydrated electrons involving $^{17}$O nuclei in the first solvation shell bias the observation towards the isotope configurations in which no $^{17}$O nuclei are present in this shell. With these assumptions, we were able to quantitatively account for the linewidths of the EPR spectra for trapped $e^-_{hyd}$, [4,5] both with and without $^{17}$O enrichment. The same DFT calculations also account for all of the important features of the ESEEM spectra, [2,6] including line widths and the presence of high frequency bands that up until now have not been explained. We believe that the residual disagreement between our calculations and experiment stems from the fact that our MQC MD calculations are based on a pseudopotential [23] that slightly overestimates the size of the solvation cavity, resulting in reduced isotropic and anisotropic hfc's as compared to the experiment.

Alternatively, the structure of the alkaline glass might be different from liquid water, resulting in tighter solvation cavities.

Our DFT calculations also yielded significant downshifts for all of the vibrational modes in the water molecules forming the solvation cavity. None of the shifts observed in our model can be accounted for by a simple electrostatic interaction with a point charge at the cavity center: instead, the observed changes in the vibrational modes result from the presence of the excess electron density in the $O\ 2p$ orbitals. The magnitude of our calculated downshifts compare favorably to those determined experimentally. [19] For electrons trapped at the surface of small water anion clusters, a similar view (red shifts originating from donor-acceptor stabilization between the unpaired electron and O-H $\sigma^*$ orbitals) has been recently suggested by Herbert and Head-Gordon. [37]

In conclusion, we believe that our hybrid DFT(CIS)-MQC MD model not only captures all of the salient features of one-electron models of the $e_{hyd}^-$, but presents a further refinement of the picture of electron hydration in general and provides, for the first time, a consistent explanation of those properties of the $e_{hyd}^-$ that cannot be addressed using one-electron models. Our DFT(CIS)-MQC MD model suggests that the traditional cavity picture of the $e_{hyd}^-$ is incomplete: the excess electron cannot be considered fully independently of the valence electrons in water molecules. Thus, we view the 'hydrated electron' as a kind of multimer radical anion [38] of water in which the electron wavefunction is shared between the cavity and the water molecules forming it. Just such a picture has been advocated by Symons [39] and, later on, by Kevan. [3]

## 6. Acknowledgement.

IAS thanks C. Elles, S. E. Bradforth, A. Boutin, J. M. Herbert, A. Khan, W. Domcke, A. Nilsson, J.-P. Renault, and D. M. Bartels for useful discussions. This work was supported by the Office of Science, Division of Chemical Sciences, US-DOE under contract No. DE-AC-02-06CH11357. BJS, WJG and REL gratefully acknowledge the support of the National Science Foundation under grant number CHE-0603776.

**Supporting Information Available:** (1) A PDF file containing Figures 1S to 16S with captions. (2) A PPT file containing SOMO maps ($\pm 0.03$ $a_0^{-3}$ contours) of hydrated electron for 50 consecutive MQC MD snapshots ($\Delta t = 100$ fs); these frames can be played as a movie of electron dynamics in liquid water using the "SlideShow" option. This material is available free of charge via the Internet at http://pubs.acs.org.

**Figure Captions.**

**Figure 1**

(a) Histograms of (i) the coordination number $n_H$ and (ii) the cluster size $n$ for the hydrated electron, $r_{cut}$=4.75 Å (in this figure and Figures 2 to 6, 8, 9, 11, and 12, the average of 1000 snapshots along the 100 ps of the MQC MD trajectory; B3LYP/6-311++G** model) and (b) the histogram for the population of *O 2p* orbitals.

**Figure 2**

The histogram of distances $r_{Xn}$ to *H* and *O* nuclei, from the center of mass of the electron *X* as defined in the MQC MD model, for $r_{cut}$=4.75 Å clusters. The nuclei are divided into two groups: "inside" and "outside," as explained in section 2.2. Solid lines are for hydrogens, gray lines are for oxygens. The mean $r_{XH}$ distance for $H_{in}$ hydrogens is 2.4 Å (see Table 1 for other mean values). The positions of $r_{in}$ and $r_{cut}$ radii are indicated by vertical lines.

**Figure 3**

(a,b) Two sequential snapshots of hydrated electron, $e_{hyd}^-$ (the time interval $\Delta t = 100$ fs). Isodensity maps of singly occupied molecular orbitals are shown for (from top to bottom) $\pm 0.02$, $\pm 0.04$, and $\pm 0.05$ $a_0^{-3}$. More such SOMO maps along the trajectory are given in the Supplement. The cross at the cavity center indicates the center of mass *X* of the electron in the MQC MD model; red is for positive, violet is for negative part of the SOMO wavefunction for the embedded water cluster anions ($r_{cut} = 4.75$ Å; (a) n=21, $\phi_{2p}^O \approx 0.16$; (b) n=22, $\phi_{2p}^O \approx 0.18$; in both cases the electron is sixfold coordinated).

**Figure 4**

(a) Histogram of the integrated negative density $\rho_-$ defined in eq. (4). The average is 0.12. (b) The solid line plotted to the left is the angle averaged SOMO density $4\pi r^2 \rho(r)$ given by eq. (5) (the continuous line, to the right is the integral of this radial density, which approaches unity for $r \to \infty$). The dashed line is a least-squares fit to this radial density using eq. (6), for $\lambda \approx 1.67$ Å.

The most probable position of the electron is at $r = 1.75$ Å. The features observed between 2.5 and 3.5 Å are from *O 2p* orbitals.

**Figure 5**

The histograms of (a) the gyration radius, $r_g$ (eq. (8)), and (b) the three semiaxes $r_a < r_b < r_c$ of the gyration tensor, eq. (7) (see the legend in the figure). SOMO density is used to calculate this tensor. (c) The histograms of the mean meridianal ($e_m$, eq. (9)) and polar ($e_p$, eq. (9)) eccentricities of the gyration ellipsoid (see the legend in the figure). Wide distribution of these eccentricities illustrates great variation in the shape of $e_{hyd}^-$.

**Figure 6**

(a) Kohn-Sham density of states (DOS) function for HOMO (SOMO) and the three lowest unoccupied molecular orbitals (MO's): LUMO, LUMO+1, and LUMO+2 (see the legend in the plot), with the maxima at -1.8, 0.34, 0.56, and 0.77 eV, respectively. The typical isodensity maps of such states are shown in Figures 6S and 7S in the supplement. These states are related to the *s*- and *p*-like states of $e_{hyd}^-$ in one-electron models. (b) The histograms of the corresponding transition energies (that exhibit the maxima at 2.05, 2.27, and 2.5 eV, respectively).

**Figure 7**

Simulated CIS($N$=10)/6-31+G* spectra for embedded anion clusters (first two solvation shells only; 100 ps trajectory, 1000 snapshots averaged). The bin width for the histograms is 0.1 eV. (a) The histogram of oscillator strengths for the three lowest energy states ('*p*-states') i, ii, and iii. (b) The overall histogram for the first 10 excited states (single excitations only). In the inset in panel (b), a histogram of the angles between the transition dipole moments $\boldsymbol{\mu}_{0i}$ for the lowest three excited states is shown. The dashed line is the experimental spectrum of $e_{hyd}^-$ in liquid water at 300 K.

**Figure 8**

Histograms for isotropic ($a^H$ and $a^O$, empty bars) and anisotropic ($B_{zz}^H$ and $B_{zz}^O$, gray curves) hyperfine coupling constants for embedded water anions (for $^1$H and $^{17}$O nuclei, respectively). (a) For hydrogens of the first solvation shell and (b) for oxygens of the first and the second solvation shells (see section 2.2 for the definition). The "outside" histogram is plotted to the top, the "inside" one is plotted to the bottom. The mean values are given in Table 1. Since the distribution functions for isotropic hfcc's are skewed, the most probable values are significantly lower than the mean ones. Observe the broad distribution of $B_{zz}^H$ for $H_a$ nuclei in (a), lower panel.

**Figure 9**

Modulo FT primary ESEEM spectra for trapped electron; the NMR frequency $\nu_D$ of the neutron in the X-band is 2.2 MHz. The first, the second, and the third harmonics of this frequency are indicated by solid vertical bars in panels (b) and (c); the dashed bold line indicates the frequency of 7.6 MHz used for rejection filtering of $^{23}Na$ modulation in the experimental spectra of ref. 6. (a,c) Simulated spectra (the matrix nuclei are not taken into account) and (b) the experimental data from ref. 6 for trapped electron in low-temperature 10 M Na$^{16}$OD:D$_2$$^{16}$O glass. In (a,c) the bold line is for the simulation that takes all deuterons in the cluster (the first and the second solvation shell), the dotted line is from all $D_a$ nuclei and the thin line is for $D_a$ nuclei with $r_{XD} < 2$ Å). In (b) the thin and the bold line correspond to the experimental spectra before and after the rejection filtering (intended to suppress the signal from matrix deuterons). Lines (i) and (ii) correspond to $\nu_{\alpha,\beta}^{\perp}$ and $\nu_{\alpha} + \nu_{\beta}$ frequencies, respectively (see Appendix A of ref. 1).  Curves (c) are the same as (a), after taking into account the distortions introduced due to the loss of the spin echo modulation pattern during the dead time of the EPR spectrometer and the broadening of the spectra due to the electron relaxation.

**Figure 10**

Positions of IR and Raman band centroids for geometry-optimized octahedral water clusters. Symbols are for water anions (Kevan's model of hydrated electron) shown in Figure 1(a) of ref. 1 (B3LYP/6-31+G** model); the lines are for neutral water clusters of the same geometry as these water anions. The centroid positions are plotted against the distance $r_{XH}$ between the center $X$ of the cluster and the six hydrogens in the HO groups pointing towards this common center (all other degrees of freedom in the water molecules were optimized). Trace (i) is for H-O-H bend (blue lines and filled circles in the lower panel), traces (ii) and (iii) are the symmetrical (black lines and open and filled triangles) and asymmetrical (red lines and open squares) O-H stretch modes (the upper panel), respectively. Filled symbols and solid lines are for IR centroids, empty symbols and dashed lines are for Raman centroids. Where the positions of IR and Raman centroids for a given geometry are very close to each other, only one set of the values is shown in the plot.

**Figure 11**

Simulated (a) IR and (b) Raman spectra (B3LYP/6-31+G** model), for (i) embedded water anion cluster ($r_{cut}$=3 Å, i.e., the first solvation shell only; black line), (ii) embedded neutral clusters ($r_{cut}$= 3.5 Å; red line), and (iii) embedded single water molecules (black line, see the legend in the upper panel). Spectra (i) and (iii) are ensemble averaged over 400 configurations, spectrum (ii) is averaged over 200 configurations. Band centroids (eq. (3)) are given in Table 2. There is a notable blue shift, relative to experiment, in all three major vibration bands. In (b), trace (iii) is scaled by a factor of four, to facilitate the comparison.

**Figure 12**

Simulated (a) IR and (b) Raman spectra without an excess electron. The solid lines show clusters with a negative point charge embedded at the position of the electron center of mass ($X$) in the MQC model, the dashed lines show embedded neutral water clusters of the same geometry. See Table 2 for positions of centroids.

**Table 1.**

Calculated parameters for the ground (*s*-) and excited (*p*-) states of 'hydrated electron' (matrix embedded water cluster anions with $r_{cut}$ = 4.75 Å).

| parameter | average | parameter | average |
|---|---|---|---|
| $M_2^H$, G$^2$ | 17.3±4.6 | $\langle r_{XH} \rangle_{in}$, Å | 2.43±0.12 |
| *iso* | 3.9±3.1 | $\langle r_{XO} \rangle_{in}$, Å | 3.42±0.11 |
| *aniso* | 13.4±3.4 | $\langle a^H \rangle_{out}$, G | 0.09±0.06 |
| $M_2^O$, G$^2$ | 6050±140 | $\langle B_{zz}^H \rangle_{out}$, G | 0.64±0.05 |
| *iso* | 6000±140 | $\langle B_{zz}^O \rangle_{out}$ G | -0.3±0.5 |
| *aniso* | 52±14 | $\langle \rho_s^H \rangle_{out}$, x10$^3$ | 7.7±2.8 |
| $\langle a^H \rangle_{in}$, G | 0.38±0.50 [d] | $\langle \rho_s^O \rangle_{out}$, x10$^3$ | -0.33±2.00 |
| $\langle a^O \rangle_{in}$, G | -15.2±2.2 | $\langle \rho_c^H \rangle_{out}$ | 0.34±0.01 |
| $\langle B_{zz}^H \rangle_{in}$, G | 3.7±0.5 | $\langle \rho_c^O \rangle_{out}$ | -0.69±0.02 |
| $\langle B_{zz}^O \rangle_{in}$, G | -2.5±0.7 | $\langle r_{XH} \rangle_{out}$ | 4.44±0.10 |
| $\langle \rho_s^H \rangle_{in}$ | 0.10±0.03 | $\langle r_{XO} \rangle_{out}$ | 4.79±0.10 |
| $\langle \rho_s^O \rangle_{in}$ | -0.042±0.010 | $\phi_{2p}^O$ | 0.17±0.02 |
| $\langle \rho_c^H \rangle_{in}$ | 0.11±0.05 | $\langle n \rangle$ | 19.7±2.00 |
| $\langle \rho_c^O \rangle_{in}$ | -0.56±0.03 | $\langle n_H \rangle$ | 5.7±1.0 |
| $\langle r_g \rangle$, Å | 2.74 [a] *(4.16)* [b] | E, eV [c] | -1.69±0.36 [a] *(0.42, 0.65, 0.86)±0.3* [b] |

(a) for the highest occupied state (the SOMO); (b) for the lowest three unoccupied states; (c) orbital energies, (d) the most probable value is -0.4±0.1 G. Standard deviations are given next to the mean values.

**Table 2**
Centroids for vibrational bands shown in Figures 11 and 12.

| band | type | a | b | c | d |
|------|------|------|------|------|------|
| Libration | *IR* | 828.5 | 925.4 | 949.6 | - |
| *450-1400 cm⁻¹* | *Raman* | 789.4 | 789.4 | 941.5 | - |
| H-O-H bend | *IR* | 1697.8 | 1732.7 | 1755 | 1747.4 |
| *1500-2000 cm⁻¹* | *Raman* | 1697.5 | 1677.3 | 1771.5 | 1754.1 |
| H-O stretch | *IR* | 2938.8 | 3018 | 2993.1 | 3127.8 |
| *2200- 4100 cm⁻¹* | *Raman* | 2952.8 | 2966.3 | 3029.4 | 3118.4 |

(a) 'hydrated electron' (embedded water anion cluster for $r_{cut}$ =3 Å); (b) the same geometry as in (a), for a neutral cluster containing a point charge at $X$; (c) for embedded neutral water cluster with $r_{cut}$ = 3.5 Å (see Figure 14S(a) for the size distribution); (d) embedded individual water molecules. In these calculations, the clusters/molecules were embedded in a matrix of point charges with $Q_H$ = +0.41.

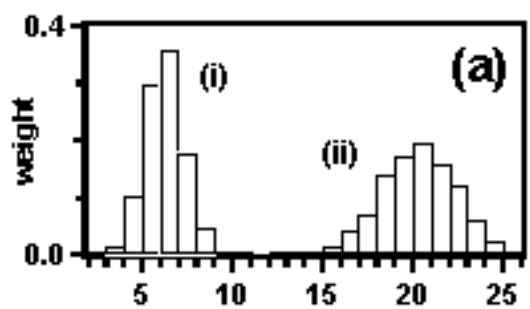

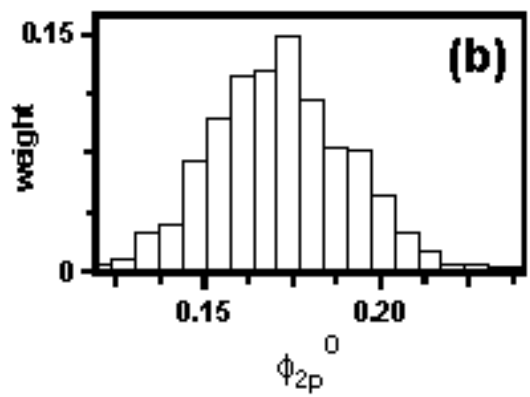

$\phi_{2p}^{\ 0}$

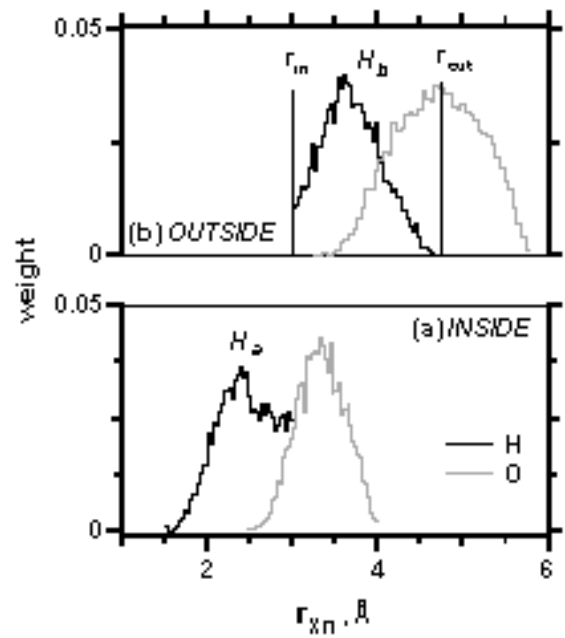

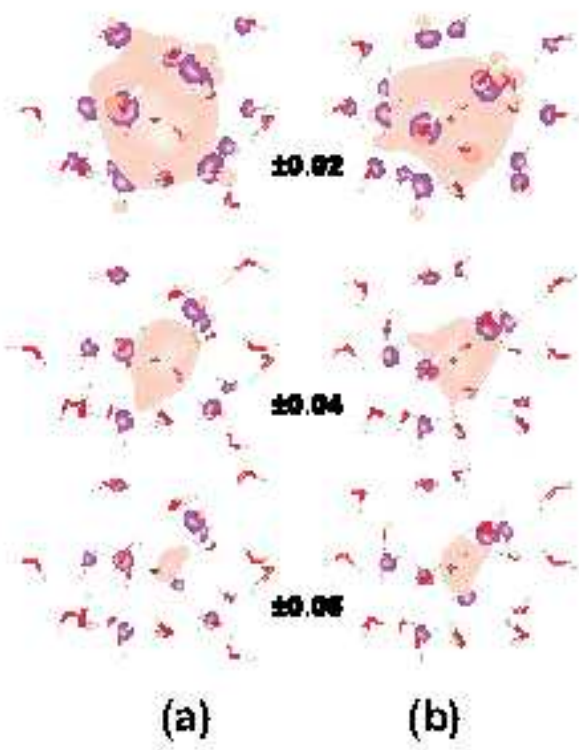

±0.02

±0.04

±0.06

(a)          (b)

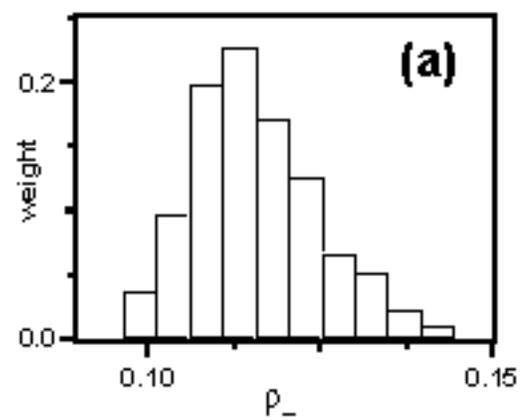

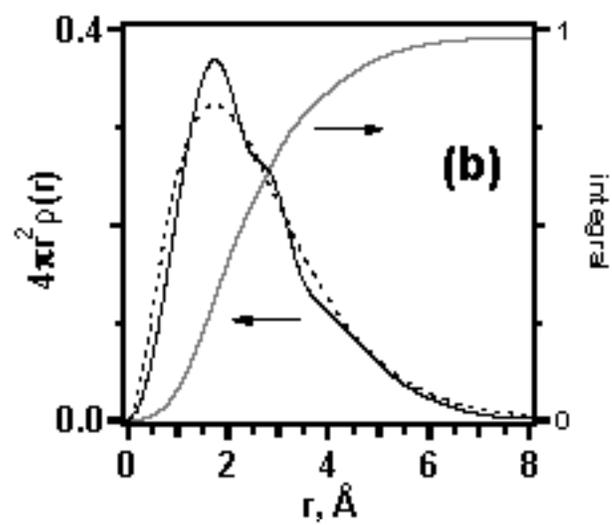

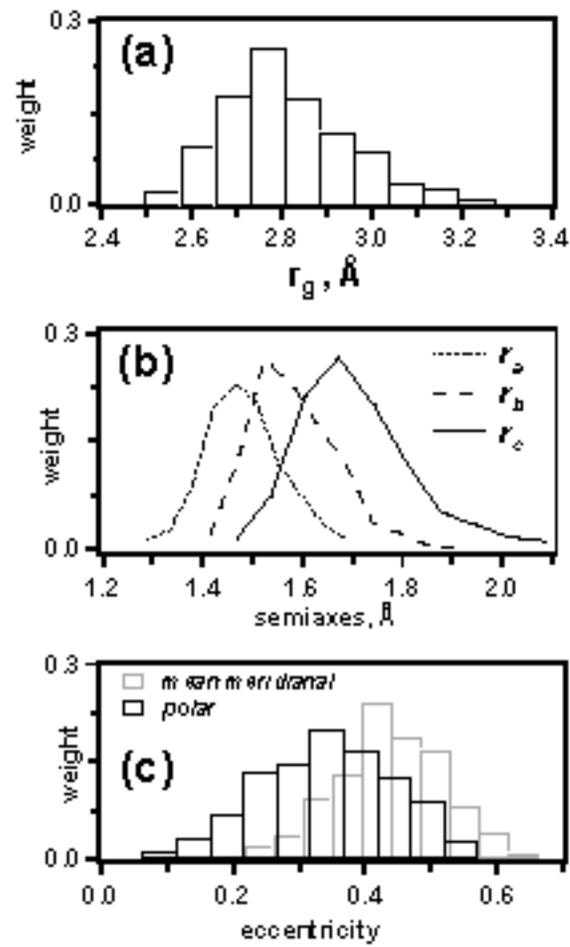

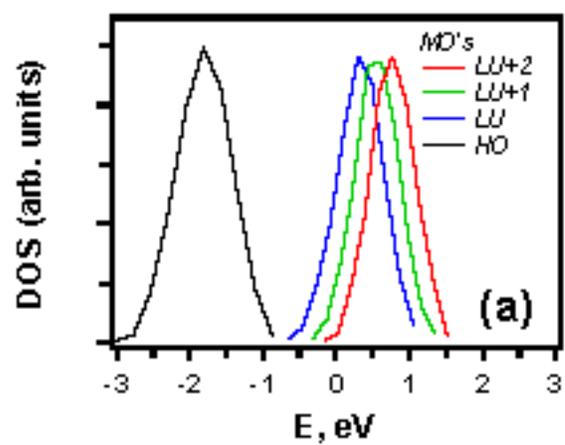

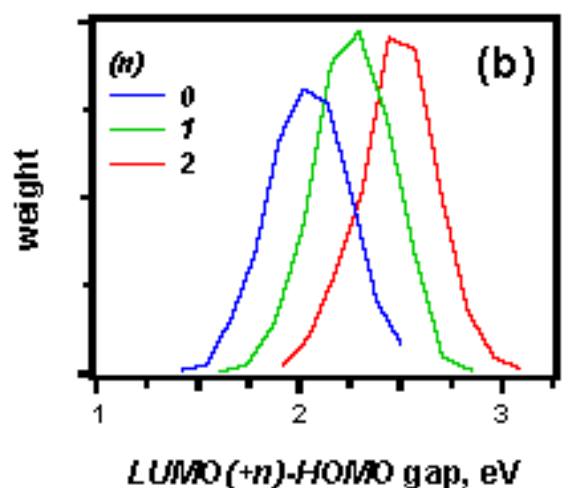

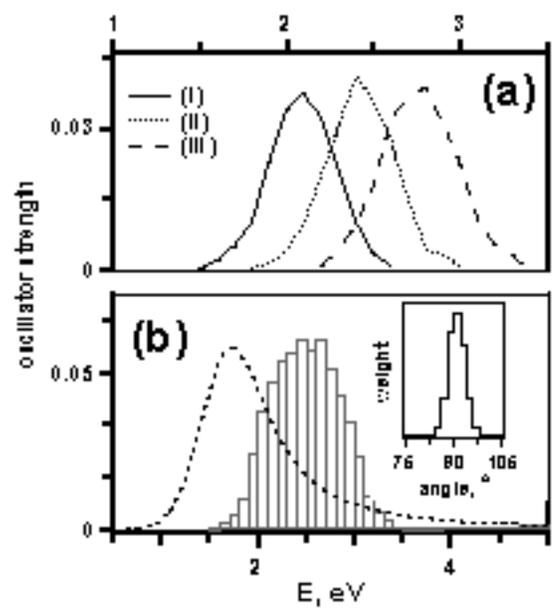

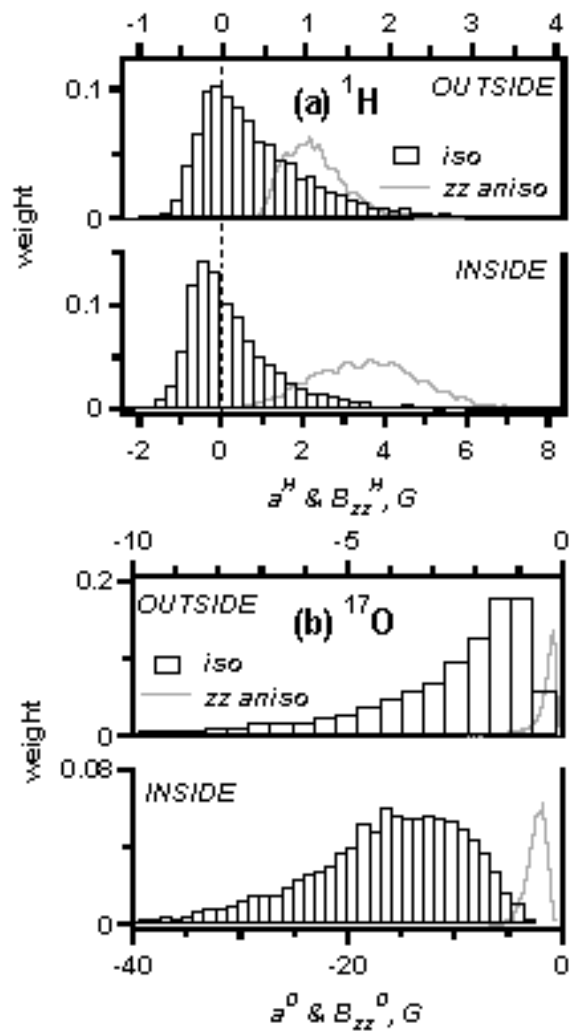

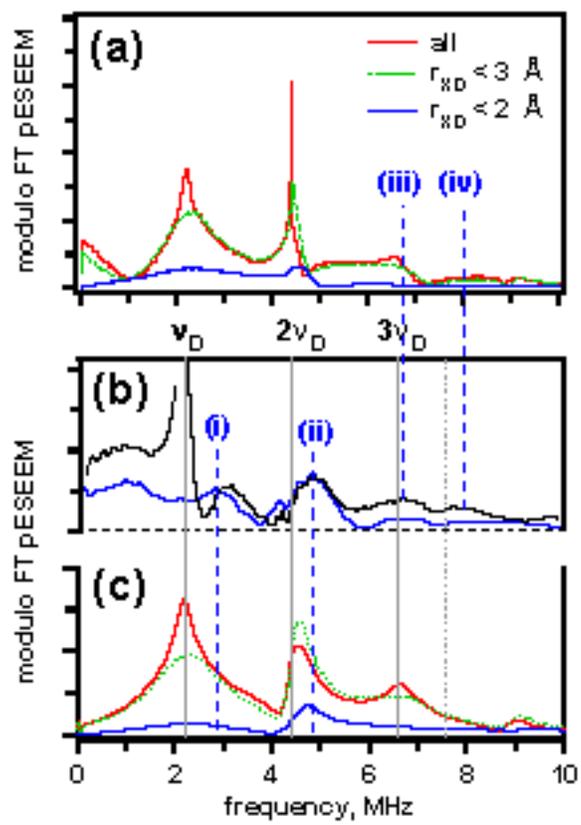

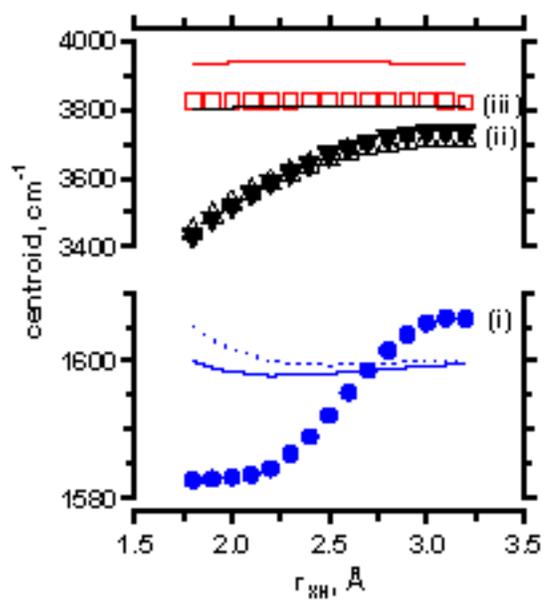

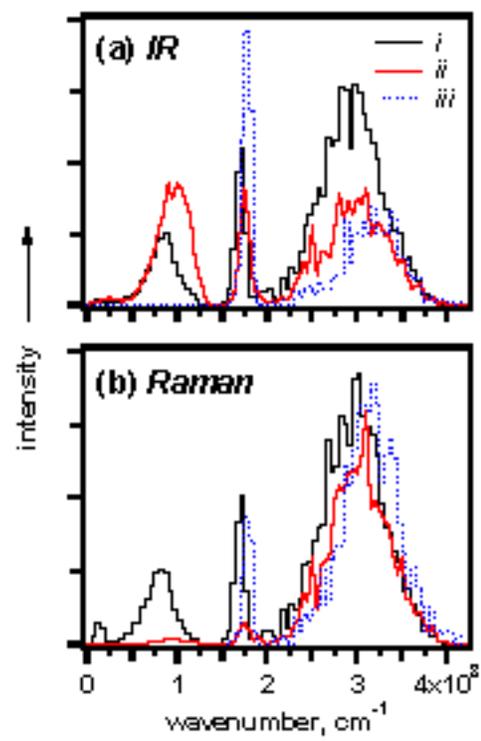

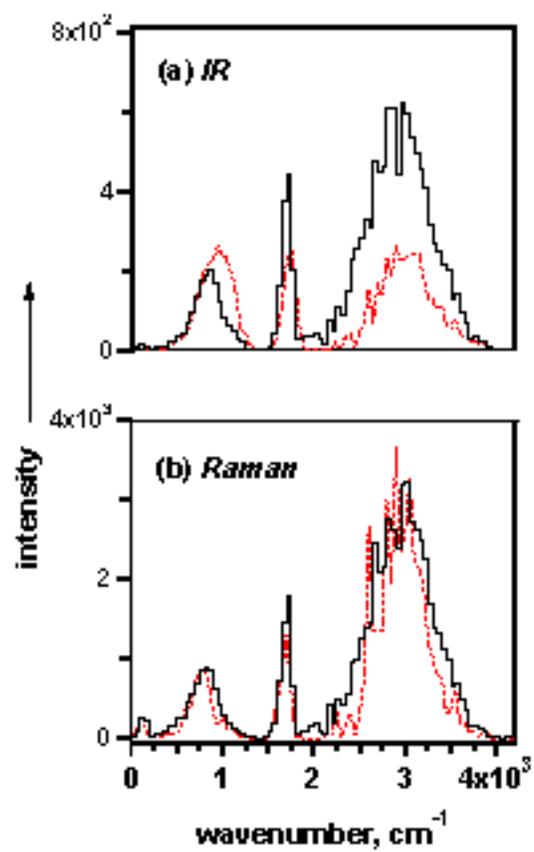

# Supporting Information.

## Figure Captions (Supplement)

### Figure 1S

(a) *Solid black line, to the left:* Pair distribution $g(r)$ of the $r_{XH}$ distances between the center of mass of the electron (MQC MD calculation, 200 snapshot average) and the protons. *Red line, to the right:* the same distribution after weighting by a factor $4\pi r^2$. Both of these distributions have the first maxima at 2.26 Å. (b) *Solid black line:* The distribution $\alpha_{XOH}$ defined as the smallest of the two XOH angles for the protons in the water molecules in the first solvation shell for which $r_{XH} < 3$ Å. The *red line* is the power-exponential fit to $\alpha^m e^{-\alpha/\alpha_0}$, where $m=1.63$ and $\alpha_0=7°$. The most probable X-O-H angle is ca. 12-14° and all of these angles are within a 60° cone, i.e. the solvating OH groups tend to point towards $X$.

### Figure 2S

(a) The correlation of the gyration radius $r_g$ of the electron, and the mean X-$H_a$ distance $\langle r_{XH} \rangle_{in}$ (open circles), for embedded water anion clusters ($r_{cut}=4.75$ Å; B3LYP/6-311++G** model, 200 snapshot average). The solid line is a linear fit: $r_g \approx 1.13 \langle r_{XH} \rangle_{in}$. In panels (b) and (c), the Kohn-Sham orbital energy for HOMO (b) and the energy gaps (c) between the HOMO and the lowest three unoccupied orbitals (see the legend for color/symbol coding) are correlated with the radius of gyration.

### Figure 3S

(a) Kohn-Sham density of states function, DOS (the occupancy number is shown) for 'hydrated electron' ($r_{cut}=4.75$ Å, which corresponds to the first two solvation shells). The arrows indicate the position of (i) HOMO (SOMO) and (ii) the three lowest unoccupied states (shown separately in Figure 6a). The red line is for occupied $\alpha$-MO's (the same spin orientation as that for the unpaired electron), the green line is for unoccupied $\alpha$-MO's; the scattered black dots are the DOS for $\beta$-MO's; the yellow line is the total DOS. (b) The same as (a), for $\alpha$-MO's in the embedded neutral water clusters ($r_{cut}=3.5$ Å; dashed blue line), and the first solvation of the hydrated electron ($r_{cut}=3$ Å clusters): the violet line is for the water anion, the yellow line is for a neutral water cluster (of the same geometry) with a negative point charge placed at the electron's center of mass ($X$).

### Figure 4S

(a) Absorption spectra of the $e_{hyd}^-$ calculated using three CIS models for the same 100 fs x 1000 snapshot MQC MD trajectory. Models (i) and (ii) are CIS($N$=10)/6-31+G* calculations (1000 snapshot average) for $r_{cut}$ of (i) 3 Å and (ii) 4.75 Å, with a ghost Cl atom at the electron's center of mass. Model (iii) is a CIS($N$=20)/6-31++G** calculation (800 snapshot average) for cutoff radius of 3 Å without the ghost atom. The dashed trace is experimental spectrum. This plot illustrates the sensitivity of the calculated CIS spectra to the details of cluster embedding and the basis. (b) The correlation of the transition energies $E_{0k}$ for the lowest ($k$ = 1,2,3) three electronically excited states of $e_{hyd}^-$ and corresponding transition moments $\mu_{0k}$, for the three lowest ('$p$-') subbands (model (iii))

**Figure 5S**

The same as Figure 7, for CIS model (iii) instead of model (ii) (see the caption to Figure 4S), (a,b) with and (c,d) without the embedding matrix of SPCf charges (800 and 200 snapshot average, respectively). In panels (b) and (d), we plotted the fits of our calculated CIS spectrum to a Gaussian-Lorentzian function that is typically used to approximate the experimental spectra of solvated electron: [32] for $\Delta E = E - E_m > 0$, the amplitude is proportional to $\left(1 + \left[\Delta E / W_L\right]^2\right)^{-1}$; for $\Delta E < 0$, it is proportional to $\exp\left(-\left[\Delta E / W_G\right]^2\right)$. For the $e_{hyd}^-$ in water at 300 K, $W_G \approx 0.42$ eV and $W_L \approx 0.49$ eV. [32] The optimum fit to our CIS spectrum in panel (b) gives $E_m \approx 2.04$ eV and 0.63 and 0.49 eV, respectively, for these two parameters.

**Figure 6S**

Typical isodensity surfaces for (a) HOMO (SOMO), (b) HOMO-1, (c,f) LUMO, (d,g) LUMO+1, and (e,h) LUMO+2 orbitals (all for the same snapshot of the 'hydrated electron'). Positive density is shown in pink, negative is shown in violet. For (a-e), the isodensity levels are $\pm0.03$ $a_0^{-3}$; for (f-h), it is $\pm0.05$ $a_0^{-3}$. The directions of the transition dipole moments are indicated by arrows. These three directions are orthogonal for these lowest unoccupied states. Here and in figures 6S to 12S, the data are for the DFT/6-311++G** model with $r_{cut}$ =4.75 Å.

**Figure 7S**

Isodensity surface for the LUMO at four isodensity levels: (a) $\pm0.01$, (b) $\pm0.02$, (c) $\pm0.03$, and (d) $\pm0.04$ $a_0^{-3}$.

**Figure 8S**

(a) As Figure 4b, for the lowest three unoccupied Kohn-Sham orbitals ('$p$-states'). The solid line drawn through the empty circles is the least squares fit to $\Psi_p(r) \propto r \exp\left(-r/\lambda\right)$ for $\lambda \approx 1.8$ Å. (b) The distributions of semiaxes of the gyration tensor for the three '$p$-states' (see the legend).

**Figure 9S**

Histograms of (a) spin and (b) charge densities for embedded water anion clusters, as determined using Mulliken population analysis, for (i) $H_{in}$ atoms that have $r_{XH} < 3$ Å, (ii) O atoms in the first solvation shell, (iii) $H_{in}$ atoms that have $r_{XH} > 3$ Å, and (iv) O atoms in the second solvation shell (see section 2.2 for the definition of these atom groupings). The spin density in the latter oxygen atoms is close to zero. In panel (b), solid lines give the charge densities on oxygen (on the left) and hydrogen (on the right) for embedded (neutral) water monomers. Both the unpaired electron and the excess charge density are limited primarily to the first solvation shell.

**Figure 10S**

The correlation plots of (a) isotropic hfcc on oxygen-17 vs. the X-O distance and (b) the zz (long axis) component of the anisotropic hfc tensor for protons (the experimental estimate is 7 G) vs. the same value estimated in point-dipole approximation (see Appendix A in ref. 1 for more detail). In (a), "inside" (open circles) and "outside" (open squares) correspond to oxygen nuclei in the first and the second solvation shell respectively (see the legend and section 2.2). The dots are values for every $^{17}O$ nucleus in a cluster, symbols are cluster average values for every snapshot. The solid line in (a) is the fit to $\propto \exp(-2r/\lambda_O)$; the optimum length parameter $\lambda_O \approx 1.59$ Å of this fit is close to the localization radius ($\lambda \approx 1.67$ Å) of the electron in the SOMO, see Figure 4b.

**Figure 11S**

Histograms of (a) $^{17}O$ and (b) $^1H$ contributions to the second moments $M_2^O$ and $M_2^H$, respectively, to the EPR spectrum of ('trapped') hydrated electron (the mean values are given in Table 1). The same calculation as in Figure 8.

**Figure 12S**

Simulated EPR spectra of $e_{hyd}^-$ in (a) $H_2^{16}O$ and (b) 1.7:1 $H_2^{16}O$:$H_2^{17}O$ solid water (the composition of the sample in the experiment of Schlick et al [5]). See Appendix A of ref. 1 for the details of the simulation procedure. The dots are the histogram of resonance offsets $\Delta B$. The red line is the convolution of this histogram with the Gaussian line broadening function (the broadening was assumed to be 1 G for (a) and 5 G for (b)); the green lines (to the right) are the first derivatives of the convoluted EPR spectra. The black curve in (a) is a Gaussian fit to the convoluted spectrum. The peak-to-peak line width $\Delta B_{pp}$ (the field interval between the points of maximum slope in the EPR spectrum or the maxima in its first derivative) is 9.1 G vs. the experimental 9.5±0.5 G (Astashkin et al. [6]). The black curve in (b) is a fit using two Gaussian functions; their derivatives ((i) and (ii)) and the sum are shown in the same plot. The broad component (i) with $\Delta B_{pp} \approx 89$ G (corresponding to $M_2 \approx 1980$ $G^2$) is from isotope configurations corresponding to at least one oxygen-17 in the first solvation shell of $e_{hyd}^-$. The narrow component (ii) with $\Delta B_{pp} \approx 23$ G (vs. experimental $18 \pm 1$ G) [5a] and $M_2 \approx 135$ G$^2$ (vs. experimental 155 G$^2$) [5a] is from isotope configurations in which the electron coupled only to $^{17}$O nuclei in the second solvation shell. The calculation does not take into account paramagnetic relaxation in the electron strongly coupled to $^{17}$O nuclei in the first solvation shell.

**Figure 13S**

The data of Figure 10 replotted vs. the total population $\phi_{2p}^{O}$ of O *2p* orbitals: (a) H-O stretch, (b) H-O-H bending modes. The downshift of the stretch mode linearly increases with the population of the frontal *2p* orbitals in the first -shell water oxygens.

**Figure 14S**

(a) A histogram of cluster size for embedded neutral water clusters ($r_{cut} = 3.5$ Å) used as reference system to produce traces (ii) in Figure 11. (b) Centroid positions of (empty circles, to the left) H-O stretch and (empty squares, to the right) H-O-H bend modes vs. the charge $Q_H$, in atomic units, on matrix protons (that are regarded as point charges; $Q_O = -2Q_H$, for neutrality). In the rest of the paper, we assumed $Q_H = 0.41$, as in the standard SPCf model. [22]

**Figure 15S**

Expanded version of Figure 12(a). The solid lines are as in Figure 12(a) (solid red line is for the full model, dashed blue line is for a point charge model). The corresponding stick spectra are shown in the same plot to illustrate the effect of binning (pink sticks are for the full model, green sticks are for the point charge model).

**Figure 16S.**

Vibrational density of states, VDOS (the density of normal modes) calculated for embedded anion and neutral (Figures 11 and 12, respectively) SPCf water clusters (solid red and dashed blue lines, respectively). Only the low energy range ($< 2000$ cm$^{-1}$ is shown). Observe the red shift of the 1000 cm$^{-1}$ (libration) and 1700 cm$^{-1}$ (H-O-H bend) bands and the prominent 100 cm$^{-1}$ (3 THz) band corresponding to vibrations of the water molecules relative to each other. This low-frequency band is barely seen in the IR spectra shown in Figures 11 and 12, as such vibrations have very low oscillator strength.

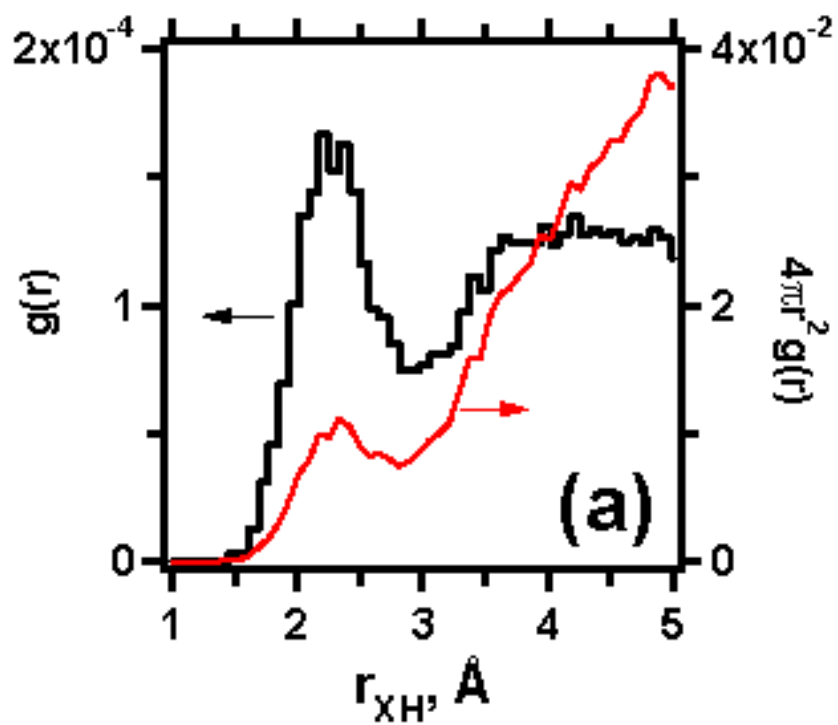

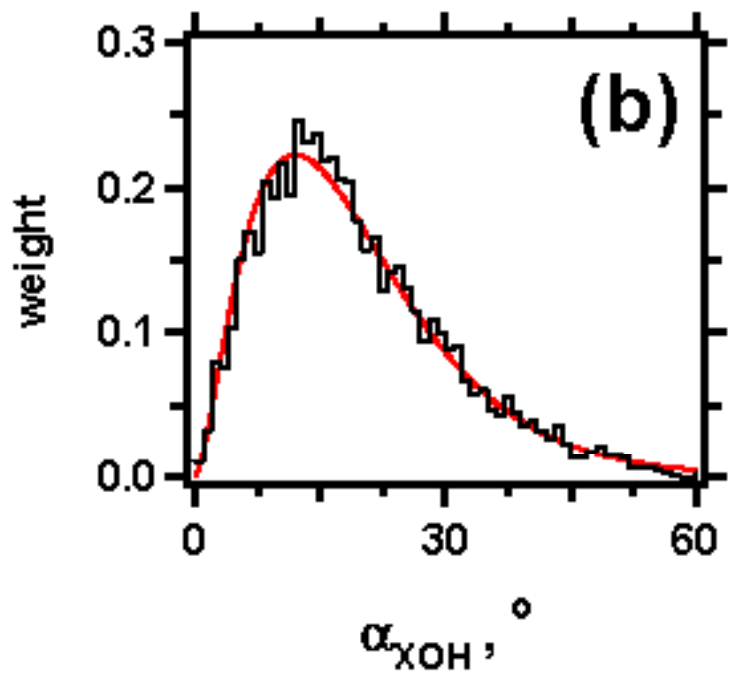

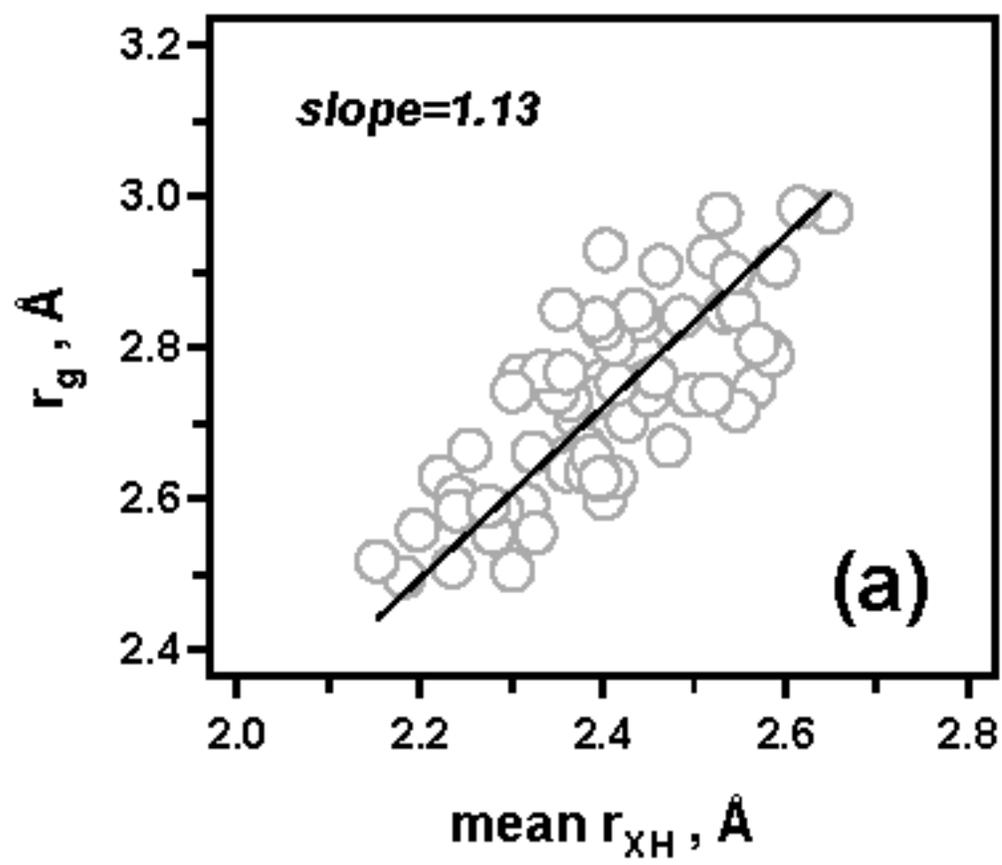

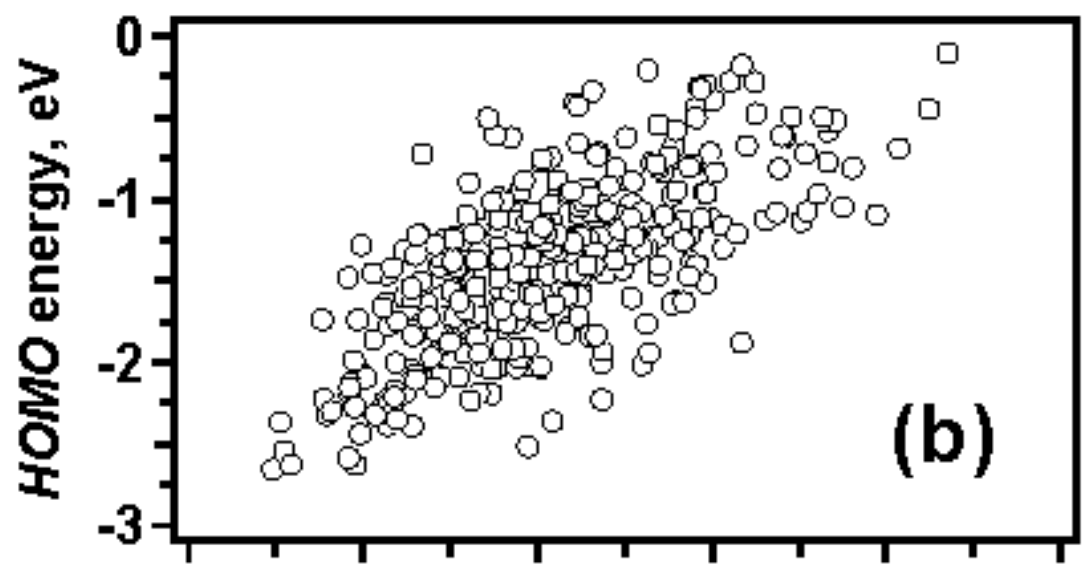

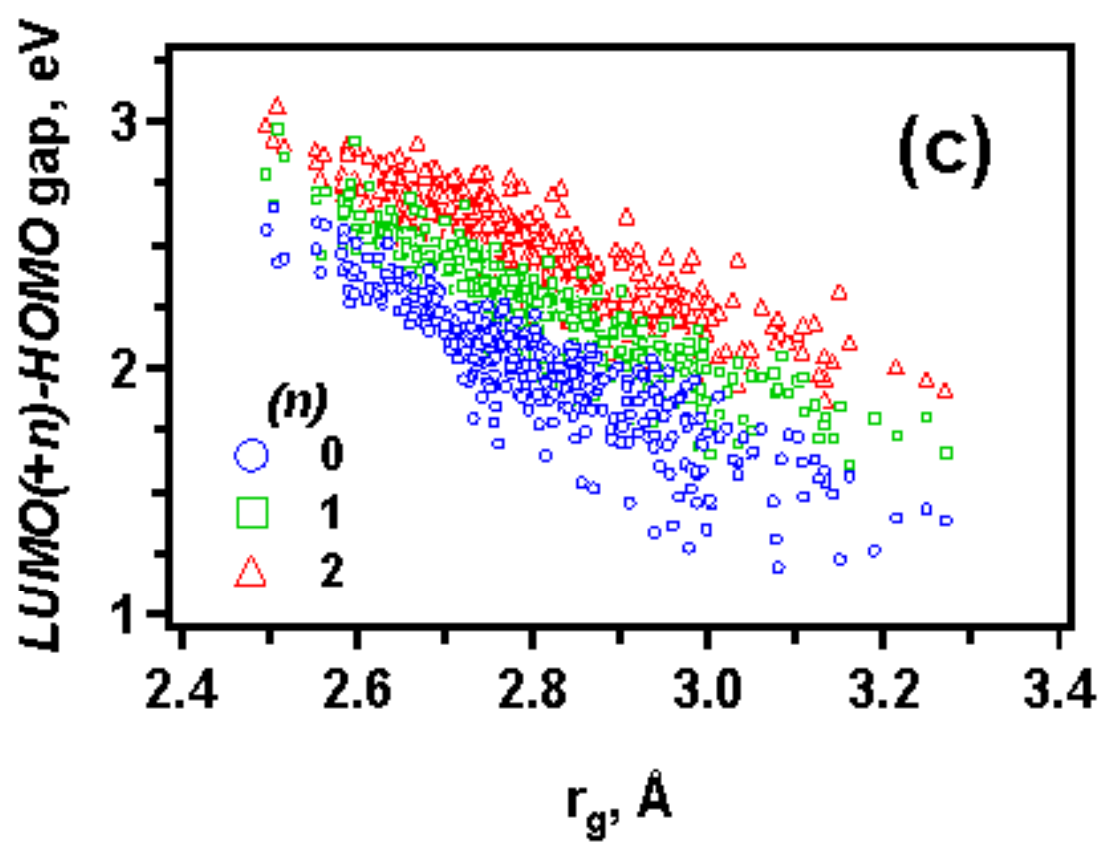

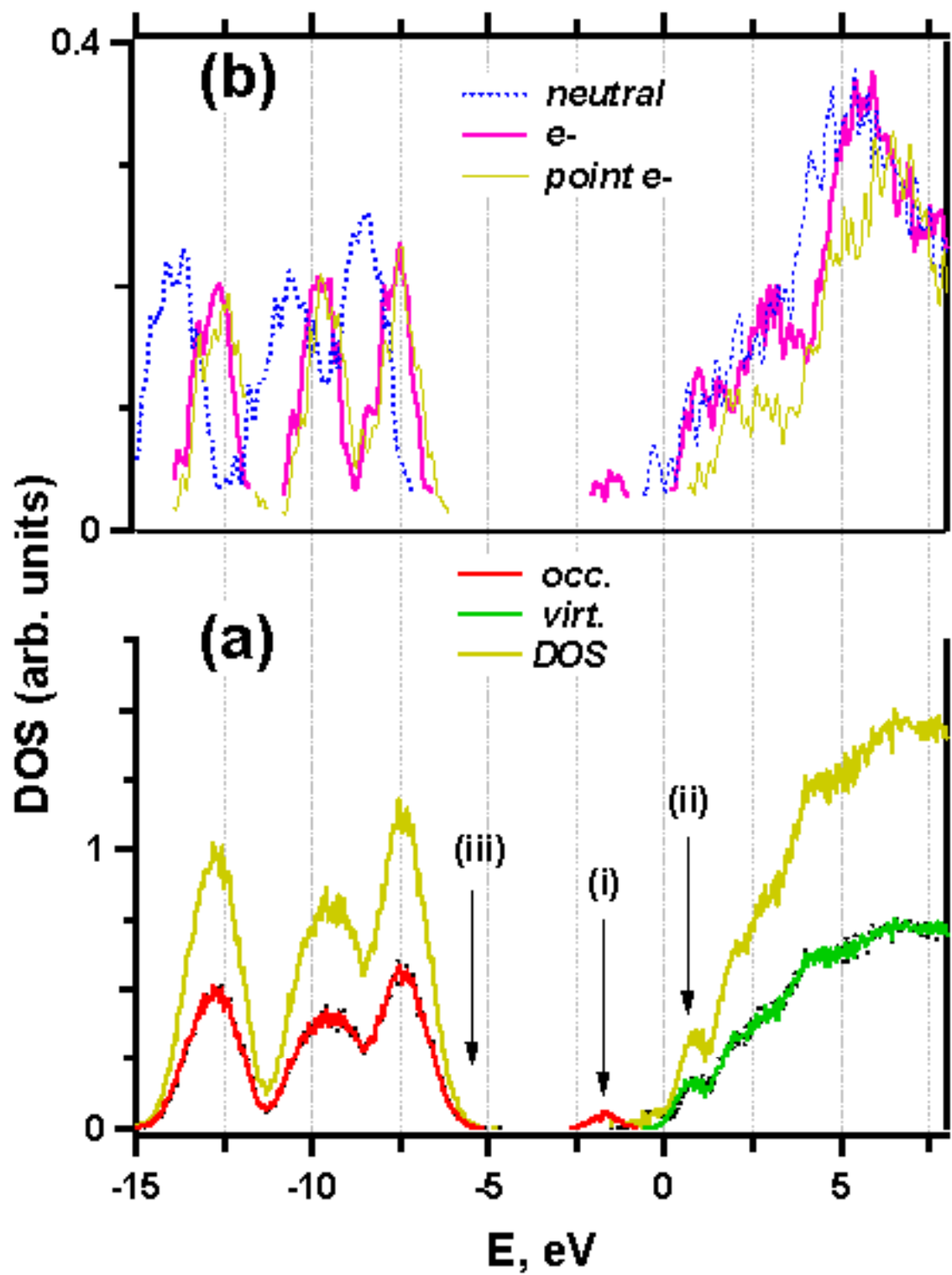

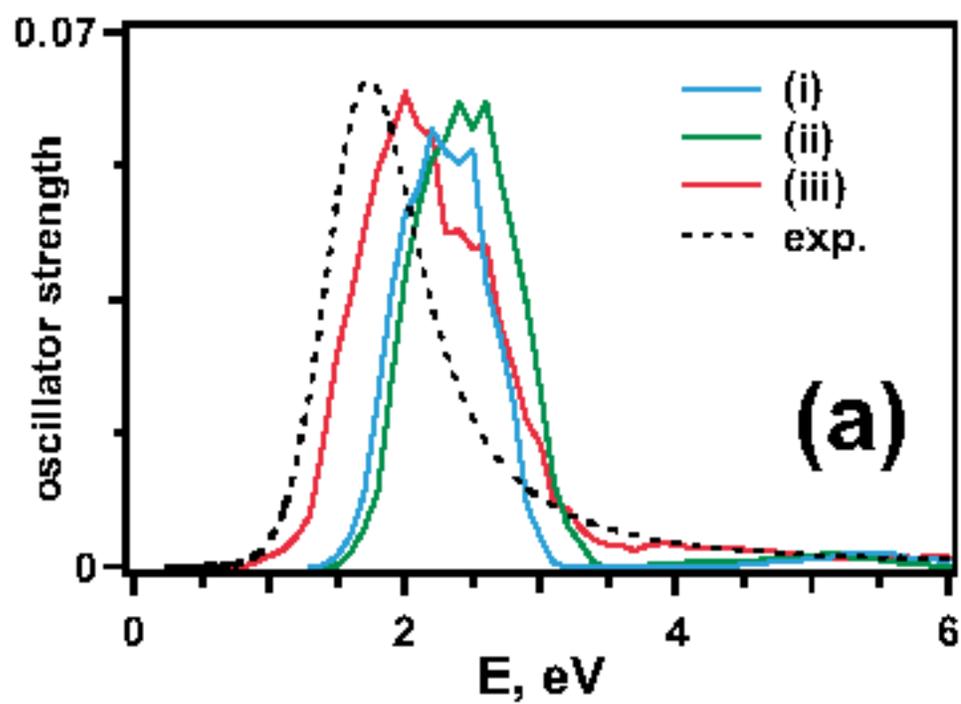

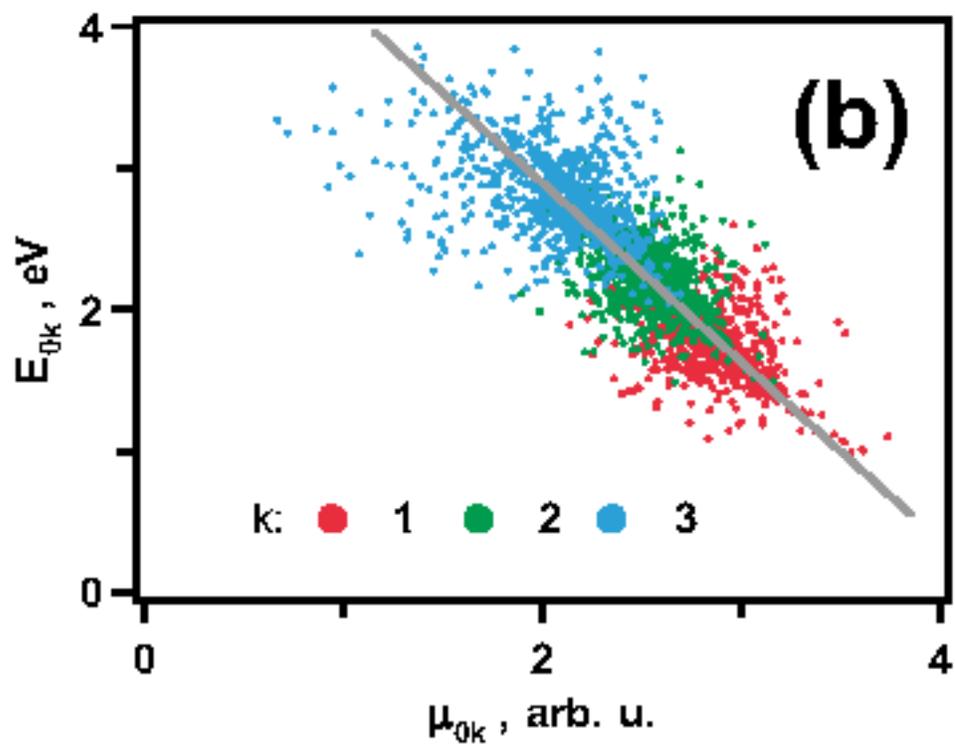

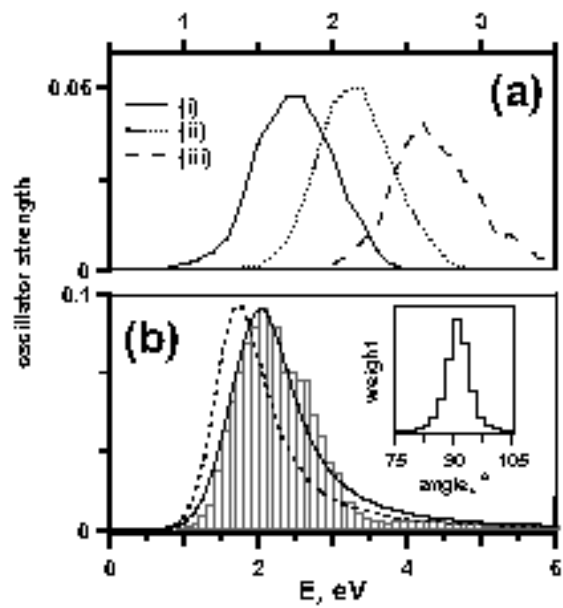

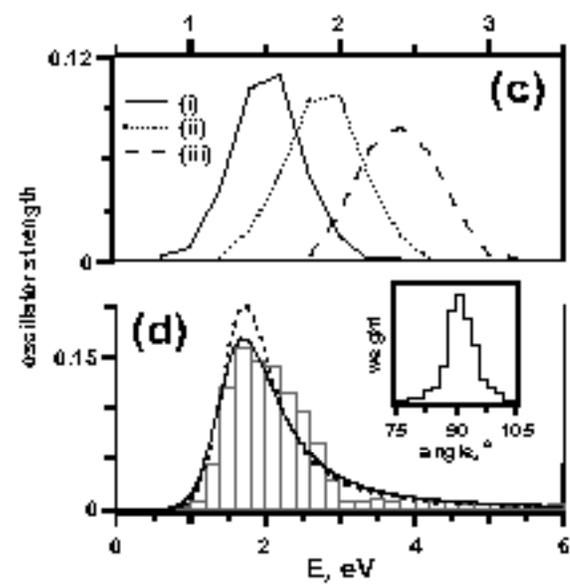

±0.03

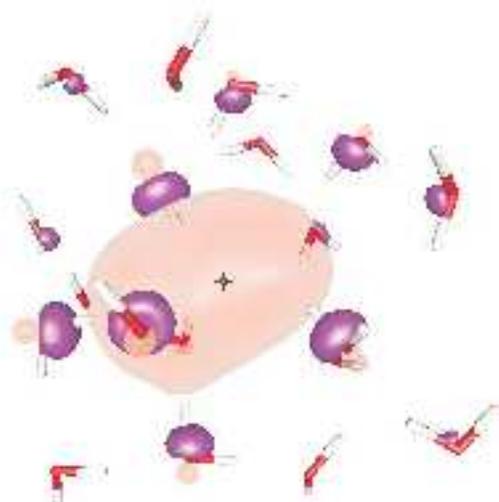

(a) HOMO

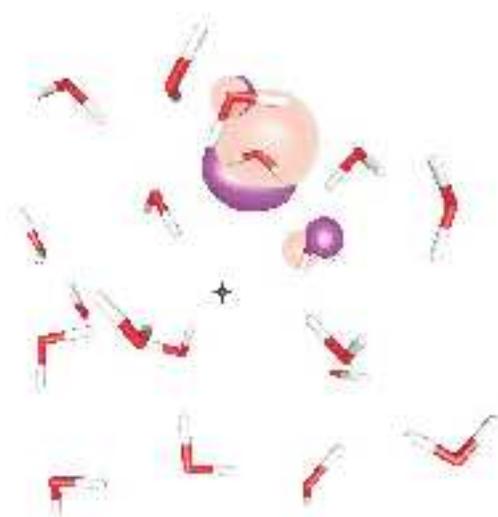

(b) HOMO-1

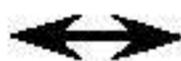

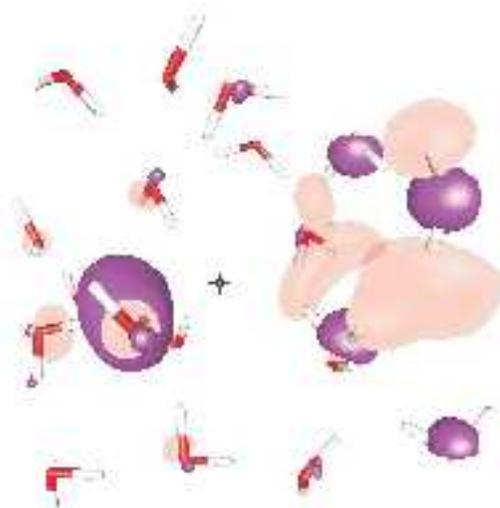

(c) LUMO

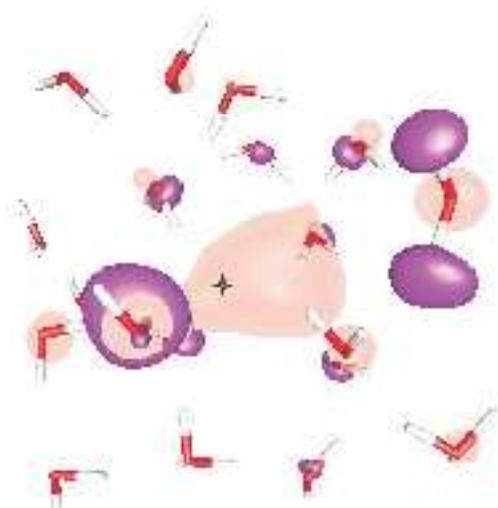

(d) LUMO+1

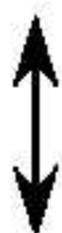

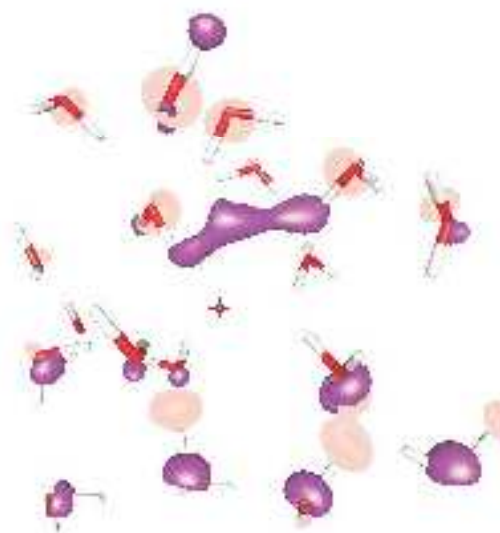

(e) LUMO+2

±0.05

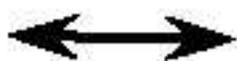

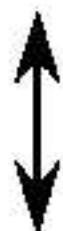

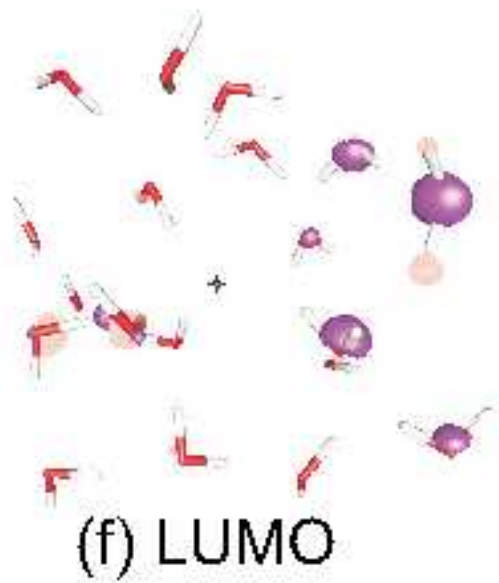

(f) LUMO

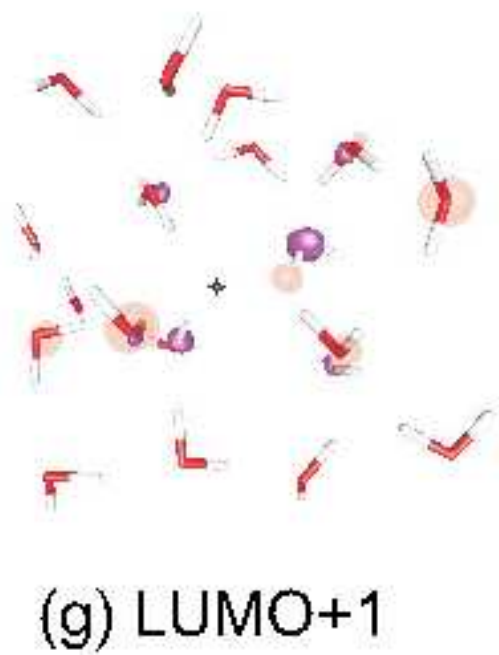

(g) LUMO+1

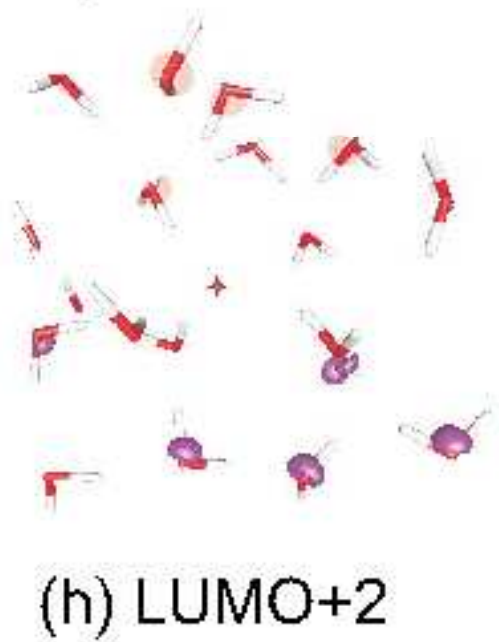

(h) LUMO+2

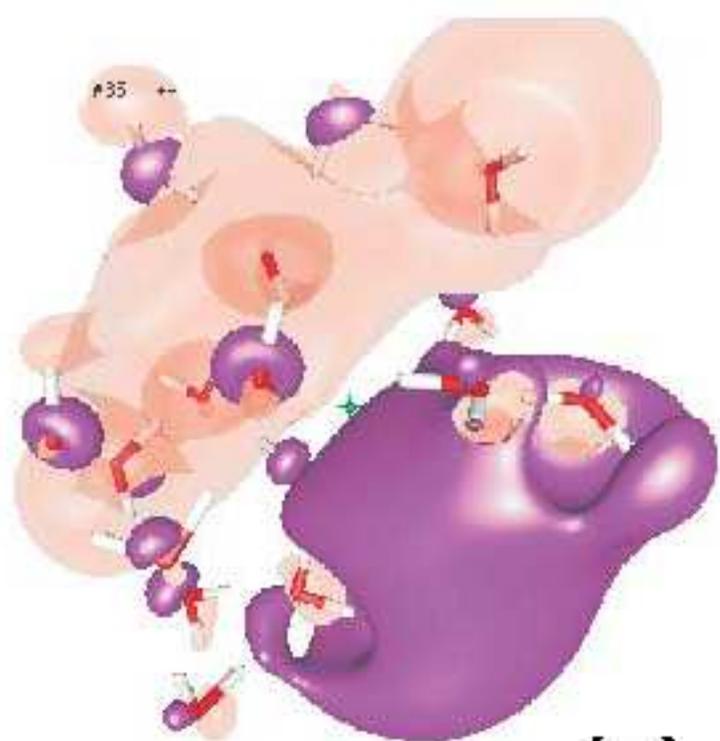
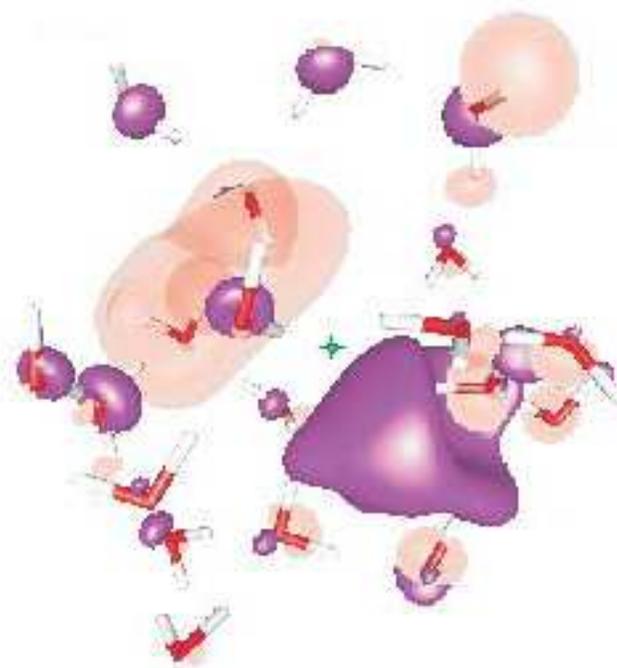

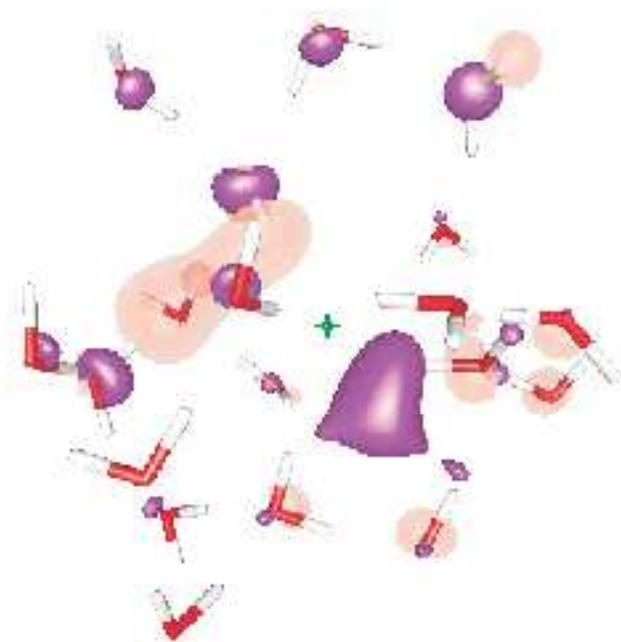
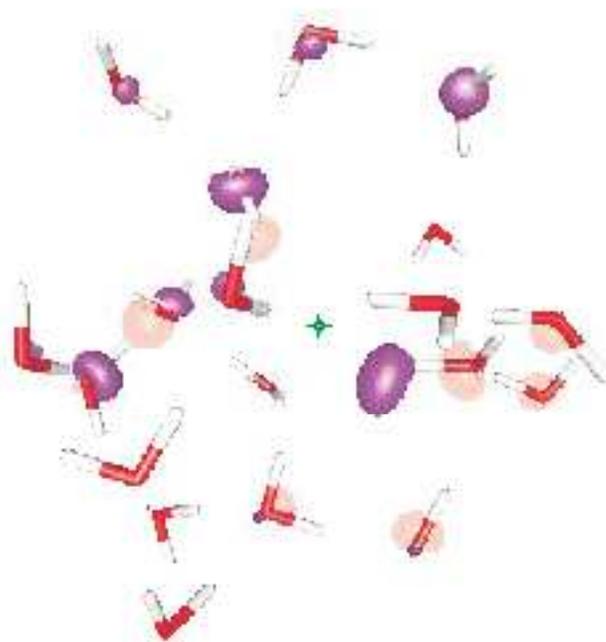

(a)  (b)

(c)  (d)

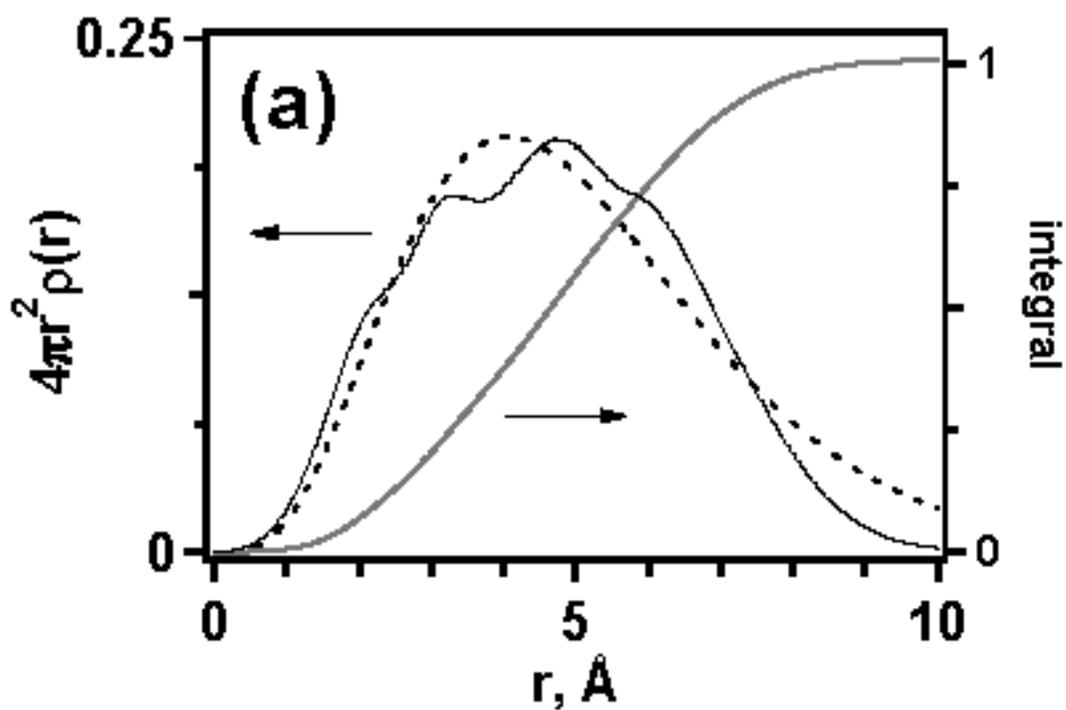

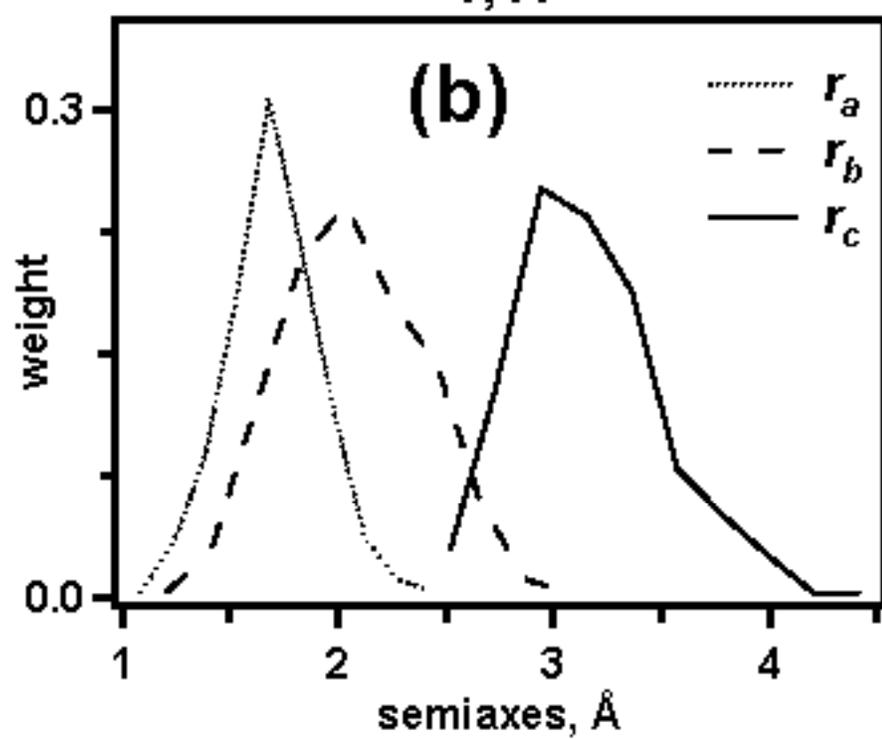

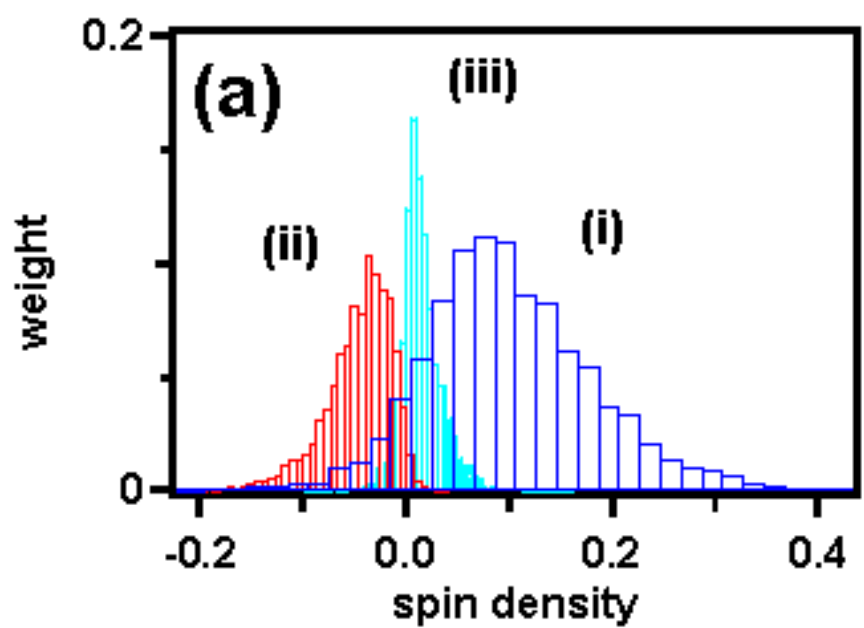

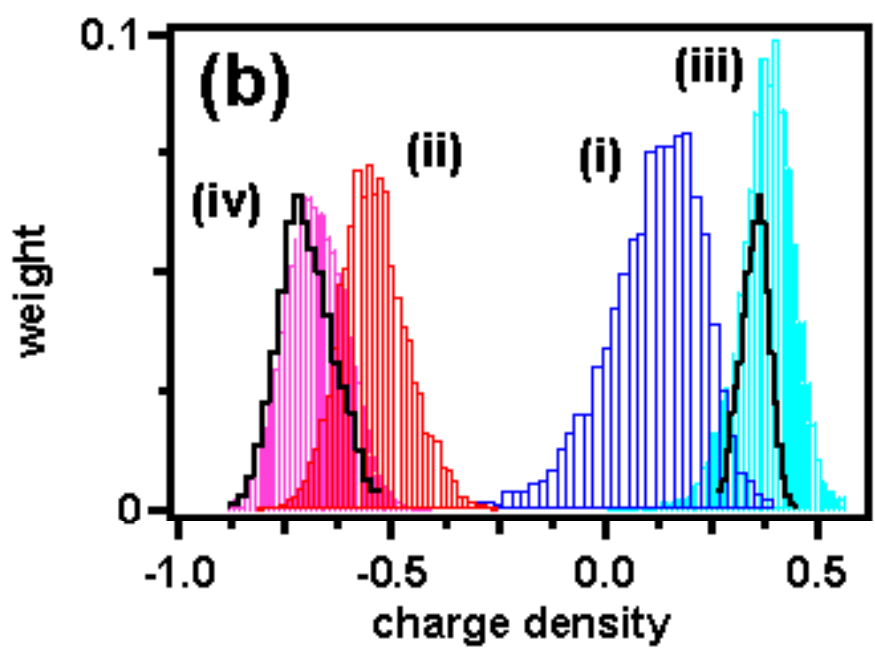

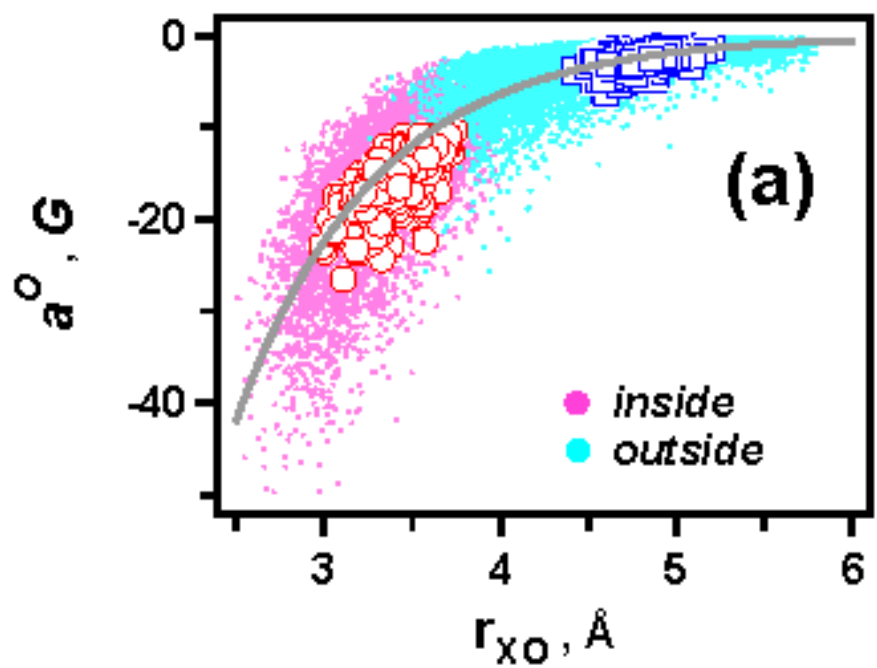

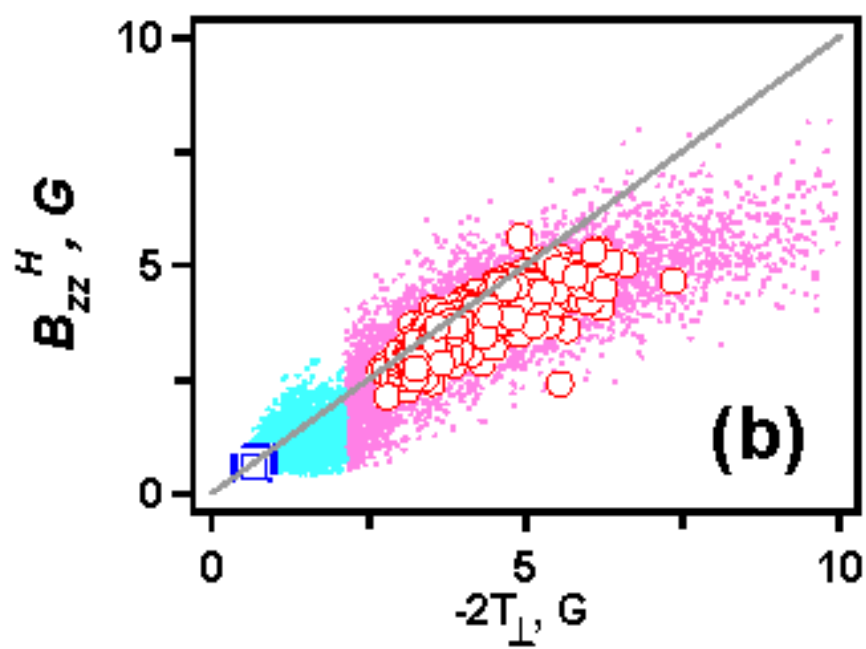

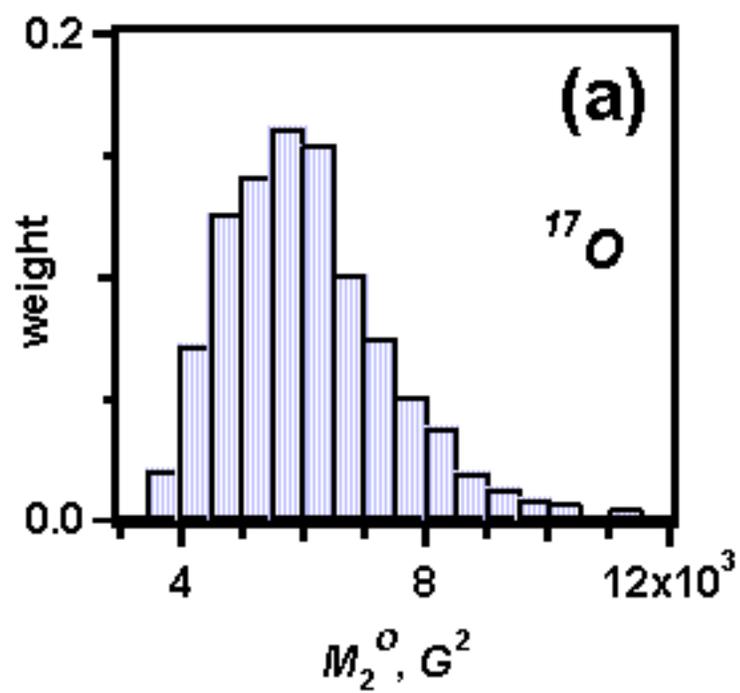

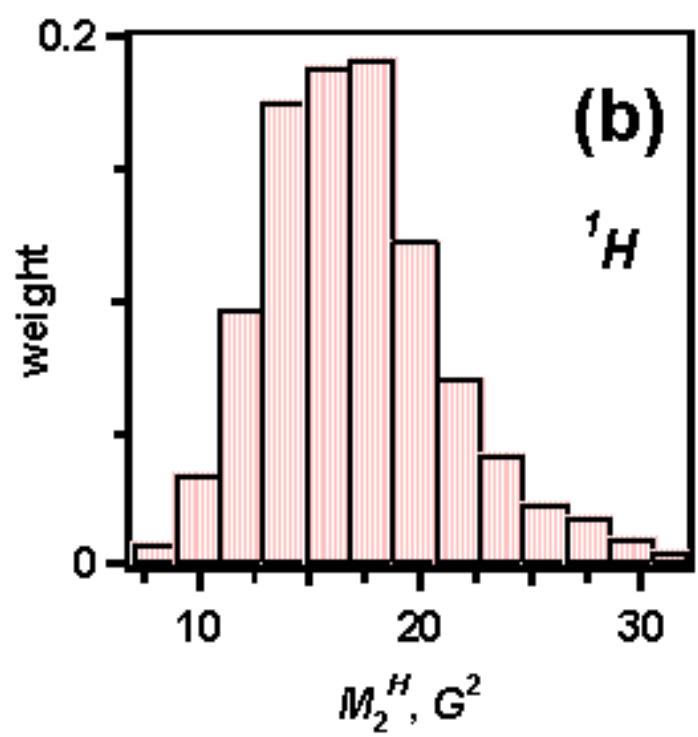

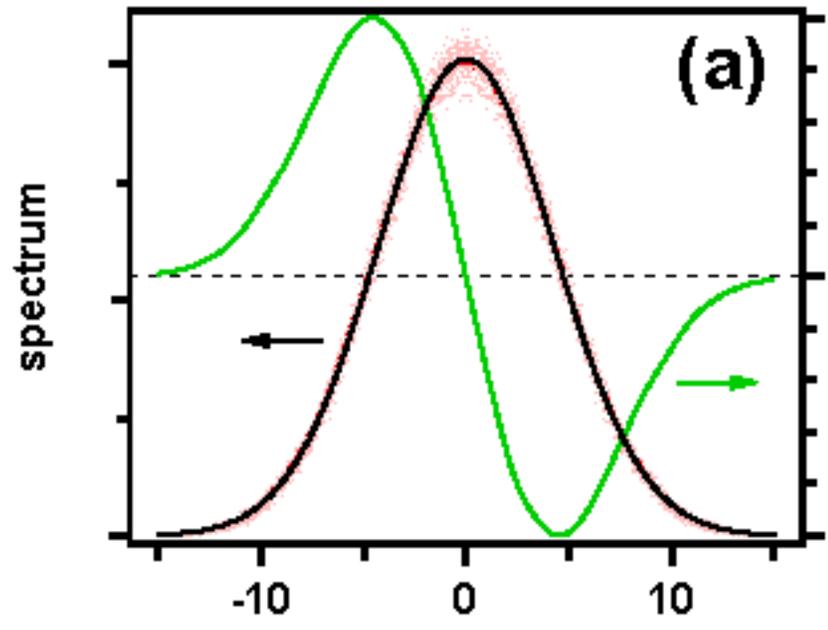

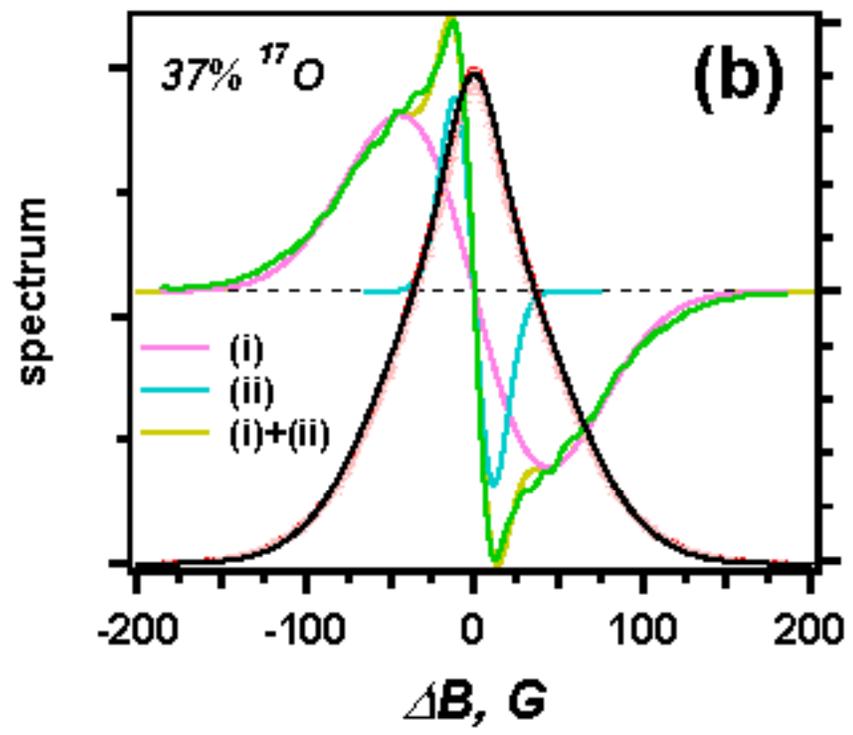

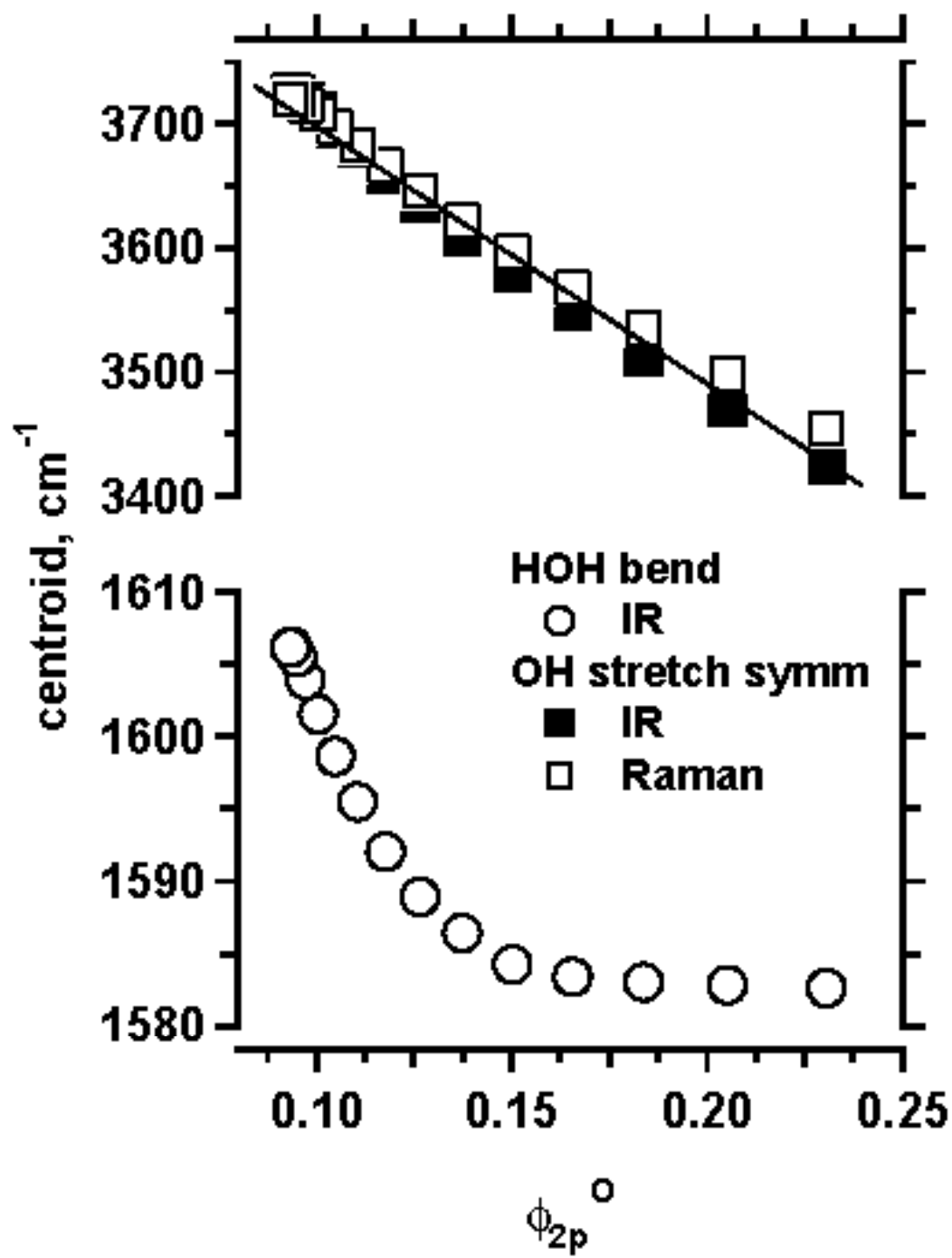

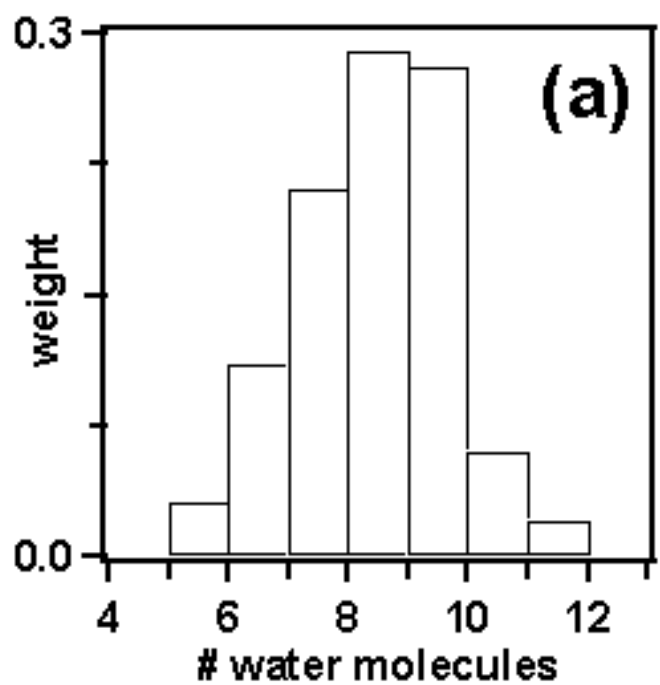

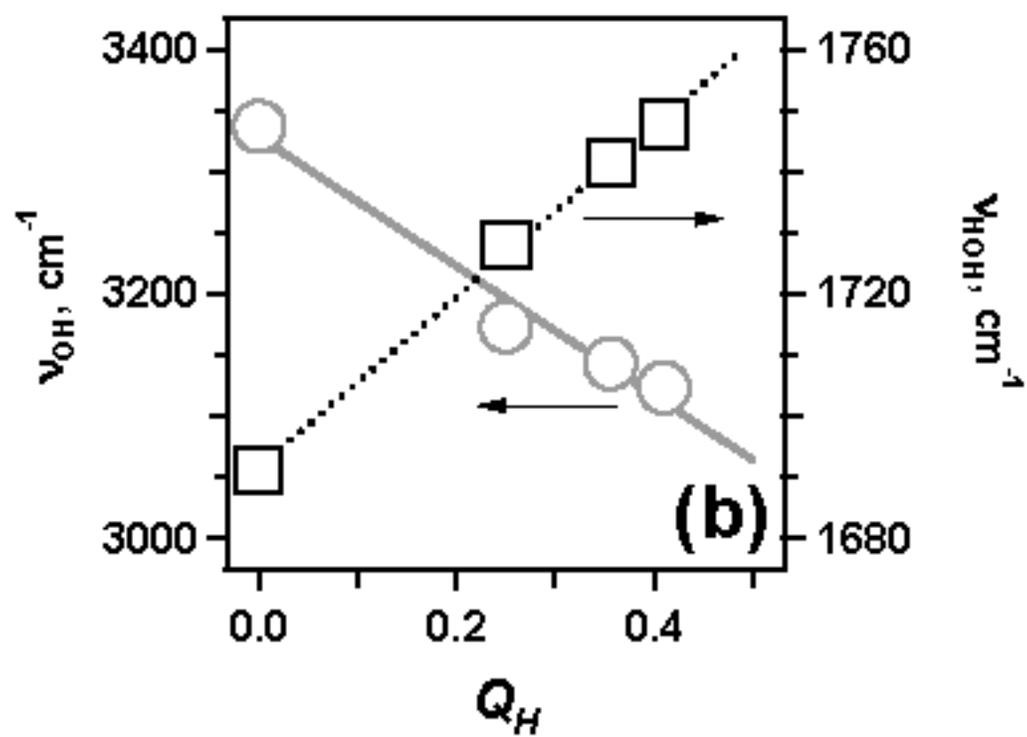

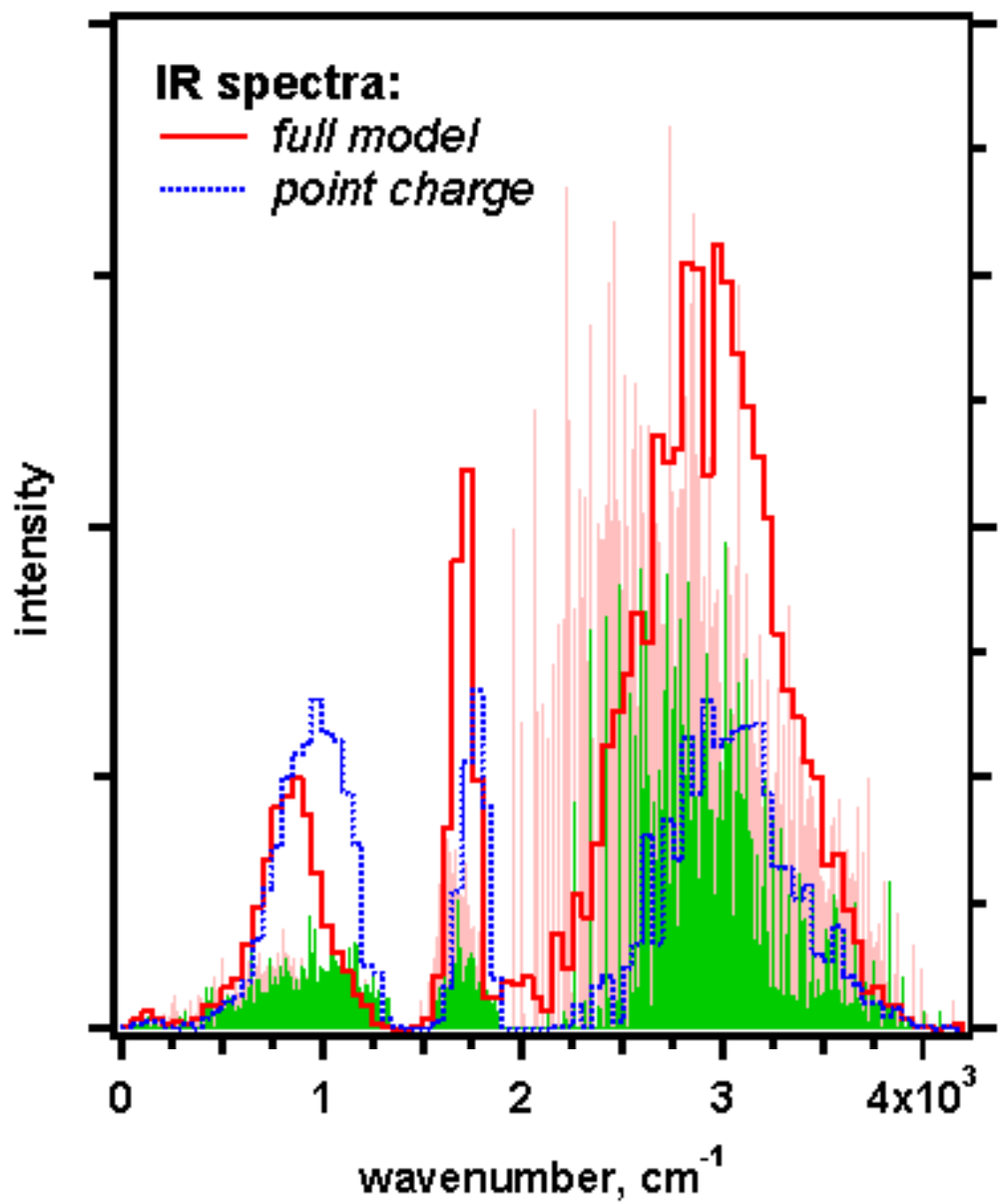

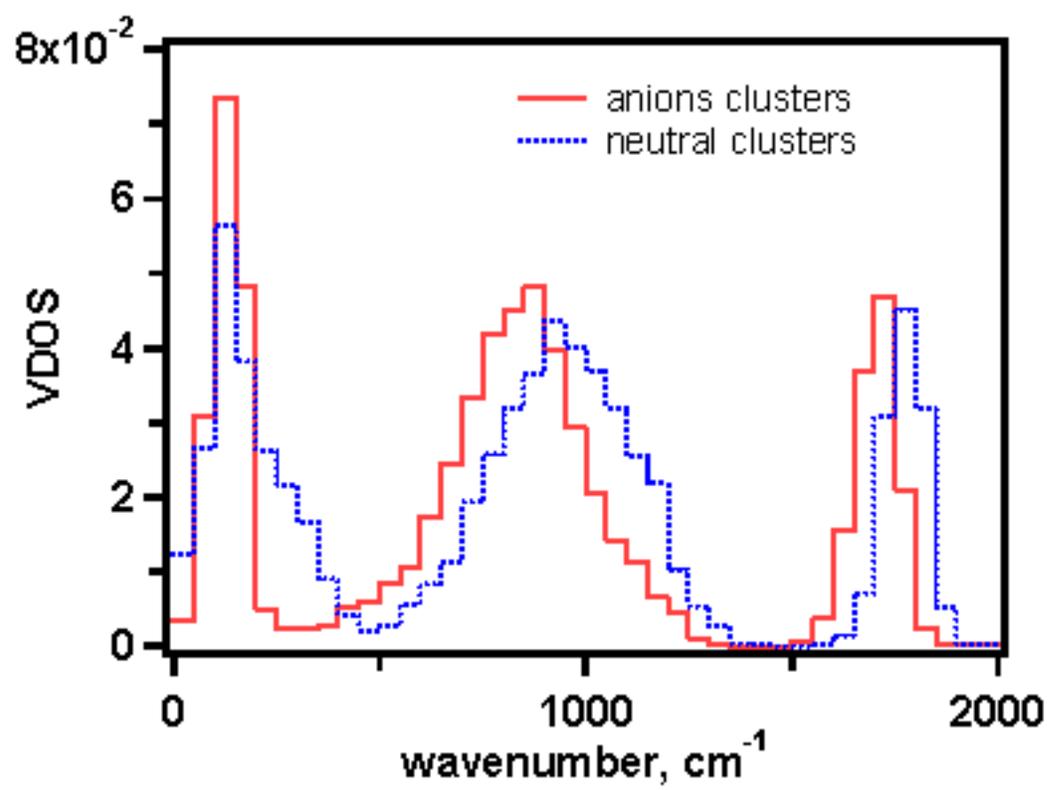